\documentclass[varenna]{cimento}

\usepackage{graphicx}  
\usepackage{longtable}
\usepackage{amssymb}
\usepackage{amsmath}
\usepackage{mathrsfs} 
\usepackage{bm}
\usepackage[utf8]{inputenc}
\usepackage{caption}
\usepackage{subcaption}
\usepackage{url}

\usepackage{makecell}
\newcommand{\ket}[1]{\mbox{\ensuremath{\vert #1 \rangle}}}
\newcommand{\beq}{\begin{equation}}
\newcommand{\eeq}{\end{equation}}

\newcommand{\Dy}{$^{161}$Dy }
\newcommand{\K}{$^{40}$K }

\title{Fermionic quantum mixtures with tunable interactions}
\author{Rudolf Grimm \atque Cosetta Baroni}

\institute{Institute for Quantum Optics and Quantum Information (IQOQI)\\ 
Austrian Academy of Sciences - Innsbruck, Austria
\vspace{3mm}\newline
Institute of Experimental Physics, University of Innsbruck - Innsbruck, Austria
}

\begin{document}
\maketitle

\begin{abstract}
The topic of the present lecture notes are two-species quantum mixtures composed of a deeply degenerate Fermi gas and a second component, the latter being fermionic or bosonic. A key ingredient is the possibility to tune the $s$\mbox{-}wave interaction between the different species by means of magnetically controlled Feshbach resonances, which allow us to investigate regimes of strong interactions. In two case studies, we review our experiments on mixtures of $^6$Li fermions with fermionic $^{40}$K or bosonic $^{41}$K atoms and on mixtures of fermionic $^{161}$Dy with $^{40}$K atoms. We cover various topics of fermionic quantum many-body physics, ranging from impurity physics and quasiparticles over phase separation to the formation of ultracold molecules and progress towards novel superfluids.

\end{abstract}

\pagebreak
\tableofcontents

\section{Introduction}

The great progress achieved in the last three decades in the field of ultracold atomic and molecular quantum matter has been accompanied by a sequence of topical Varenna summer schools \cite{Varenna1991book, Varenna1998book, Varenna2006book, Varenna2014book}. The Course~CCXI, which took place in June 2022, centered around the exciting topic of quantum mixtures. The manifold experimental possibilities to combine components of different properties or
to create systems with new properties have opened up many new research directions.

In these lecture notes, we focus on the physics of {\em degenerate fermions in quantum mixtures} and, in particular, on many-body phenomena emerging in the case of {\em resonant interspecies interactions}. After setting the stage with a brief introduction into the general ideas of the field of ultracold mixtures (Secs.~\ref{ssec:mixtures} and \ref{ssec:speciesmix}) and a short account of interaction tuning via Feshbach resonances (Sec.~\ref{sec:FR}), we discuss two special cases of fermionic mixtures (Li-K and Dy-K) implemented in our laboratories. In Sec.~\ref{sec:LiK}, we present a collection of our results from experiments on mixtures of fermionic (or bosonic) potassium atoms with fermionic lithium. In this system, studies on quasiparticles (polarons) in a Fermi sea is a central research topic. In Sec.~\ref{sec:DyK}, we present a new mixture that combines fermionic dysprosium with fermionic potassium with the motivation to create novel types of superfluids. 



\subsection{Quantum mixtures of three kinds}
\label{ssec:mixtures}

\begin{table}[t]
\caption{Comparison of the main properties of spin mixtures, isotope mixtures, and species mixtures (see discussion in the text). Entries ``no'' mean strictly impossible, while ``(no)'' means not possible or realistic in practice.}
\label{tab:mixtures}
\includegraphics[width=1.0\linewidth]{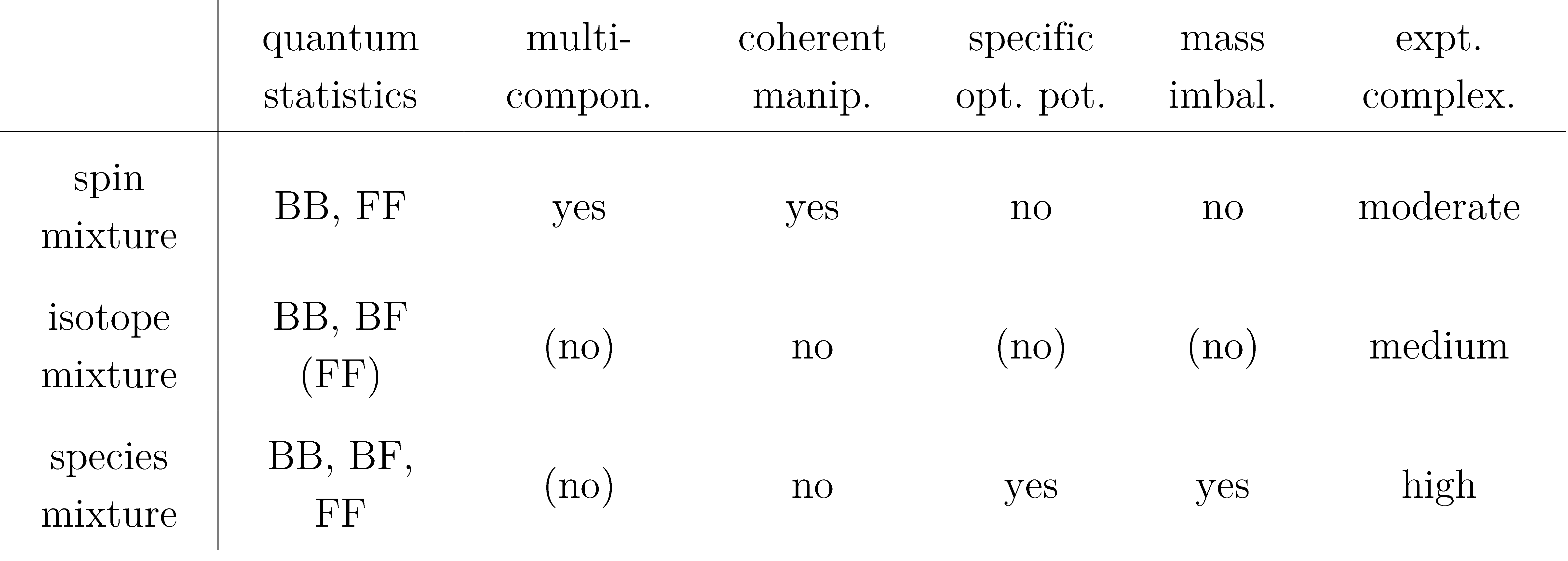}
\end{table}

Ultracold mixtures can be divided into three different classes, depending on what makes their components distinct: The general properties of {spin mixtures}, {isotope mixtures}, and {species mixtures} are summarized in Tab.~\ref{tab:mixtures} and discussed in the following.

{\em Spin mixtures} are based on atoms of the same species, but with population in different spin or hyperfine states (see Refs.~\cite{Myatt1997pot, Stenger1998sdi} for early experiments with BECs). Such mixtures are either bosonic or fermionic~\footnote{A notable exception are mixtures where a part of the fermionic atoms pair up to form composite bosons \cite{Shin2008roa}.},
and we refer to them as Bose-Bose (BB) or Fermi-Fermi (FF) mixtures. Multi-component systems can be realized by populating more than two spin states. The spin composition can be manipulated by radio-frequency and optical fields, in principle in a fully coherent way. In an optical dipole trap~\cite{Grimm2000odt}, all atoms in a spin mixture usually experience the same optical potential~\footnote{This is true as long as non-scalar terms in the atoms's ground-state dynamic polarizability stay negligibly small. This is usually the case for far-detuned trap light.}. This keeps the spin as a completely free degree of freedom, which is essential for many applications. As a drawback for certain applications, however, the possibilities to realize state-specific optical potentials are very limited. A mixture of different spin states does not offer any mass imbalance. On the practical side, the experiments require only moderate efforts, since a single species has to be prepared and the spin manipulation can be done with radio-frequency techniques.


{\em Isotope mixtures} can be realized with a variety of chemical elements commonly used in quantum-gas experiments. The main benefit unfolds if both bosonic and fermionic isotopes are present~\cite{Truscott2001oof, Schreck2001qbe}. In this case, Fermi-Bose (FB) mixtures can be realized with the same atomic species. With isotope shifts of resonance lines being usually much smaller than the optical detuning of the trapping light, the two components experience essentially the same optical potentials. Moreover, since the mass ratio of the two isotopes stays close to one, mass imbalance is practically absent. The realization of such mixtures in the lab requires additional laser frequencies to cover the isotope shifts, but the overall apparative concept (e.g.\ atomic beam source) stays the same. Multi-component isotope mixtures would be realizable in principle, but with limited practical interest.

{\em Species mixtures} offer maximum flexibility as they allow experimentalists to combine atoms with vastly different properties. All cases of combined quantum statistics (BB, FF, FB) can be realized. Since different species exhibit largely different wavelengths for optical dipole trapping, species-specific potentials are rather the rule than the exception. Mass imbalance can be realized in a wide range (up to an extreme mass ratio of nearly 30), which is of particular importance for few-body phenomena~\cite{Petrov2022bmf, Zaccanti2022mif}. The price one has to pay for this great flexibility is a relatively large experimental complexity, regarding laser systems and the whole apparative concept. 

More complex mixtures can be created in the laboratory in various hybrid ways. Notable examples are an FF species mixture coexisting with a Bose-Einstein condensate (BEC) \cite{Taglieber2008qdt}, a strongly interacting FB mixture immersed in a Fermi sea \cite{Wu2011sii}, and superfluid FF spin mixtures coupled to a BEC of another isotope~\cite{Ferrierbarbut2014amo, Salomon2022xxx} or coupled to a BEC of another species~\cite{Yao2016ooc, Roy2017tem}.
There are no limits to the imagination, but the key question for starting new experiments is whether a new combination will open up the possibility to enter uncharted physical terrain.

We finally note that mixtures play a particularly important role in experiments in which fermionic atoms are cooled to quantum degeneracy. Since elastic collisions of identical fermions freeze out at ultralow temperatures~\footnote{A notable exception are dipolar fermions, see Sec.~\ref{ssec:DyKprepare}.}, efficient evaporative and/or sympathetic cooling needs spin mixtures \cite{DeMarco1999oof, Ohara2002ooa}, isotope mixtures \cite{Truscott2001oof, Schreck2001qbe}, or species mixtures \cite{Hadzibabic2002tsm, Roati2002fbq}.



\subsection{Species mixtures: brief history}
\label{ssec:speciesmix}

Experiments on mixtures of laser-cooled atomic species date back to the 1990's, when first two-species magneto-optical traps were realized for Na-K~\cite{Santos1995sto}, Na-Rb~\cite{Telles1999icc}, Na-Cs~\cite{Shaffer1999tli}, and Li-Cs~\cite{Schloder1999cic}. The interest was driven by curiosity about the properties of more complex cold atomic systems and by understanding the relevant quantum-collisional processes~\cite{Weiner1999eat}. In these early days, new applications like sympathetic cooling~\cite{Engler1998otf} and the creation of heteronuclear molecules~\cite{Stwalley1999pou} were already at the horizon, and these prospects were driving further exciting developments.

In the early 2000's, species mixtures entered the realm of degenerate quantum gases and opened up new experimental possibilities. The power of sympathetic cooling was demonstrated by cooling $^{41}$K to BEC using Rb atoms~\cite{Modugno2001bec}, and by the efficient production of degenerate Fermi gases of $^6$Li by cooling with $^{23}$Na atoms~\cite{Hadzibabic2002tsm} and of $^{40}$K atoms by cooling with $^{87}$Rb~\cite{Roati2002fbq, Goldwin2004mot}. The experiments on double-degenerate quantum gases then quickly advanced and novel collective phenomena were observed, like the collapse of a degenerate Fermi gas induced by the presence of a BEC \cite{Modugno2002coa, Ospelkaus2006idd}.

The next milestone in the field was to gain control of the 
interspecies interaction by Feshbach resonances~\cite{Chin2010fri}, which were observed in Fermi-Bose mixtures of $^6$Li and $^{23}$Na~\cite{Stan2004oof} and of $^{40}$K and $^{87}$Rb~\cite{Inouye2004ooh, Ferlaino2006fso}. 
By magnetic tuning of the interspecies interaction in Fermi-Bose mixtures of $^{40}$K and $^{87}$Rb, the controlled collapse of the mixture was demonstrated for attractive interaction along with the onset of phase separation in the repulsive regime~\cite{Zaccanti2006cot, Ospelkaus2006toh}.

Soon another important milestone was reached by exploiting Feshbach resonances for the association of weakly bound heteronuclear molecules. Such Feshbach molecules were created with K-Rb Fermi-Bose mixtures in
optical lattices~\cite{Ospelkaus2006uhm} and in the macroscopic confining potential of an optical dipole trap \cite{Zirbel2008hmi}.
These pioneering experiments opened up the gate into the world of ultracold quantum gases of polar molecules, which has now become an extremely active research field~\cite{Carr2009cau, Jin2012itu, Bohn2017cmp, Softley2023cau}.


\section{Feshbach resonances and interaction tuning}
\label{sec:FR}

The control of interactions is a key component in many experiments on ultracold quantum gases. It is accomplished by magnetically tuned resonances, called \textit{Feshbach resonances} or \textit{Fano-Feshbach resonances}. The resonance arises as a result of the coupling of a colliding atom pair (open channel) to a molecular state (closed channel). Therefore, Feshbach resonances are inherently connected to near-threshold molecular physics. The physics of Feshbach resonances is reviewed extensively in Ref.~\cite{Chin2010fri}. Here we summarize their main properties and introduce some basic definitions as used in these lecture notes. 

\subsection{Magnetic control of the scattering length}
\label{ssec:FR1}

The low-energy scattering behavior near a single, isolated resonance (which does not overlap with other resonances) can be described by the magnetic-field dependent $s$-wave scattering length
\begin{equation}
    a(B) = a_{\rm bg} \left( 1- \frac{\Delta}{B-B_0} \right) \, .
    \label{eq:FR}
\end{equation}
Here, $B_0$ represents the resonance center, where the scattering length diverges, and $a_{\rm bg}$ is the background scattering length, determined by the open channel. The width $\Delta$ is defined as the distance between the resonance center at $B_0$ and the zero crossing at $B_0 +\Delta$.

In some of our work, we use an alternative notation based on a \textit{strength parameter} defined as $A = \Delta a_{\rm bg}/a_0$, with unit of magnetic field. Here $a_0$ is Bohr's radius. Close to the pole, where a small $a_{\rm bg}$ can be neglected, the scattering length is then obtained as
\begin{equation}
    a(B) = - \frac{A}{B-B_0} \, a_0 \,.
    \label{eq:FRA}
\end{equation}

\subsection{Broad and narrow resonances}
\label{ssec:FR2}

For a complete characterization of an $s$-wave Feshbach resonance, a further parameter is required. It can be conveniently expressed in terms of a characteristic length scale
\begin{equation}
    R^* = \frac{\hbar^2}{2 m_{\rm r} \, a_{\rm bg} \, \delta \mu \, \Delta}
    = \frac{\hbar^2}{2 m_{\rm r} \, a_0 \, \delta \mu \, A} \, ,
\label{eq:Rstar}
\end{equation}
where $m_{\rm r}$ denotes the reduced mass and $\delta \mu$ represents the differential magnetic moment between the molecular state in the closed channel and the atom pair in the open channel.
The \textit{range parameter} $R^*$, introduced by D.~Petrov in Ref.~\cite{Petrov2003tbp} and used in our work, is closely related to the textbook definition of the \textit{effective range} $r_0$ \cite{Blatt1949oti, Bethe1949tot}. On resonance, where $a \rightarrow \pm \infty$, the two quantities are related by $R^*= -r_0/2$. The range parameter facilitates a classification in terms of {\em narrow} and {\em broad} resonances (or closed- and open-channel dominated resonances, respectively~\cite{Chin2010fri}). If $R^*$ is small compared with the physical range of the molecular interaction (van-der-Waals length scale $R_{\rm vdW}$), the scattering problem simplifies to the one in a single channel. In this case, universal behavior is obtained across the whole resonance, which means that the behavior is governed by the scattering length as a single parameter. Extreme examples of such broad, open-channel dominated resonances can be found in Cs \cite{Berninger2013frw} and $^6$Li \cite{Bartenstein2005pdo}. For narrow resonances ($R^* \gg R_{\rm vdW}$), which are found in many systems, such single-parameter universality is restricted to the narrow range where $|a| \gg R^*$.

\subsection{Molecular state}
\label{ssec:FR3}

For large positive scattering lengths ($a \gg |a_{\rm bg}|, R_{\rm vdW}$), a weakly bound molecular state exists, the binding energy of which can be modeled \cite{Petrov2003tbp} as
\begin{equation}
    E_{\rm b} = \frac{\hbar^2}{8 (R^*)^2 m_{\rm r}}
    \left(
    \sqrt{1-\frac{4 R^* (B-B_0)}{a_0 A}} -1
    \right)^2 \, .
\label{eq:Eb}
\end{equation}
In the universal range ($a \gg R^*$), this reduces to the simple universal expression $E_{\rm b} = \hbar^2/(2 m_{\rm r} a^2)$.

The range parameter $R^*$ 
also characterizes the momentum-dependence of a scattering event with relative momentum of $\hbar k$ and energy $\hbar^2 k^2/(2m_{\rm r})$. While, for $kR^*\ll1$, the momentum dependence remains (negligibly) weak, it can profoundly modify the scattering process at $kR^*\gg1$. For a more detailed discussion and an illustrative example, see the contribution of M.~Zaccanti to these proceedings.


\section{Mixtures of Li and K: Impurities in a Fermi sea}
\label{sec:LiK}

The chemical elements Li and K are natural candidates for the experimental realization of fermionic mixtures, being the only two alkali metals that offer fermionic isotopes. Consequently, early experiments on Fermi-Fermi mixtures carried out by us \cite{Wille2008eau, Spiegelhalder2009cso, Spiegelhalder2010aop} and other groups \cite{Taglieber2008qdt, Wu2011sii, Voigt2009uhf, Tiecke2010bfr} focused on the mixture of $^6$Li and $^{40}$K, while double-species combinations involving non-alkali atoms were launched later \cite{Hara2011qdm, Ravensbergen2018poa, Ciamei2022ddf, Schafer2023oof}. At that time, spin mixtures of fermionic atoms (in single-species experiments) had already led to spectacular achievements \cite{Varenna2006book}, and the new species-mixture experiments were driven by the intriguing prospects to introduce mass imbalance and species-selective manipulation~\cite{LeBlanc2007sso, Catani2009eei} to explore new phenomena of few- and many-body interactions.

Introducing the new Li-K mixture, the first challenge, besides developing the cooling and trapping techniques for combining the different species \cite{Spiegelhalder2010aop}, was to understand the interaction properties along with the magnetic tunability of the system. This was accomplished in joint experimental and theoretical efforts by Feshbach spectroscopy \cite{Wille2008eau}. This finally led to a complete understanding of the two-body interaction properties \cite{Naik2011fri} and provided crucial insights to develop further experimental strategies. It turned out that Feshbach resonances in the lowest spin channel, the only one that is immune against two-body decay, are very narrow. Higher spin channels offer broader resonances, but at the price of inelastic two-body losses. This does not support experiments that require collisional stability over long times. Our investigations therefore focused on other topics of interest related to strong interactions near Feshbach resonances, like the fast hydrodynamic expansion of the mixture \cite{Trenkwalder2011heo} or few-body interactions \cite{Spiegelhalder2009cso,Jag2014ooa,Jag2016lof}. 

Over the past decade, the physics of quantum impurities immersed in a Fermi sea under conditions of strong interactions, most prominently the `Fermi polaron', has become our main research topic. While our earlier experiments \cite{Kohstall2012mac,Cetina2015doi, Cetina2016umb} focused on fermionic impurity atoms in a Fermi sea (FF mixtures), we more recently extended our experiments to Fermi-Bose systems (FB mixtures) \cite{Lous2017toa, Lous2018pti, Fritsche2021sab, Baroni2023mib} to explore the rich physics of bosonic impurities. Experimentally, this turned  out to be rather straightforward by changing the impurity isotope from fermionic $^{40}$K to bosonic $^{41}$K. On one hand, this choice of different isotopes allows for a direct comparison between fermionic and non-degenerate bosonic impurities \cite{Baroni2023mib} and, on the other hand, a Bose-Einstein condensate (BEC) of impurity atoms can be seen as a realization of a  mesoscopic impurity, which introduces intriguing static \cite{Lous2018pti} and dynamic \cite{Huang2019bmo} interaction phenomena.

In this Section, we review our main results on impurity physics with Li-K mixtures. We first summarize the main properties of such mixtures and some basic experimental procedures (Sec.~\ref{ssec:LiKbasics}), before discussing Fermi polarons (Sec.~\ref{ssec:polarons}) and mesoscopic, i.e.\ condensed impurities (Sec.~\ref{ssec:meso}) in more detail.

\subsection{Experimental basics}
\label{ssec:LiKbasics}

Here we present an overview of the basic conditions and procedures of our experiments on Li-K mixtures, which have been developed since the late 2000's. We focus on the main, conceptually important points.
For details on the preparation procedure of the mixture by laser cooling and subsequent evaporative cooling, we refer the reader to the PhD theses of C.~Baroni (2023), I.~Fritsche (2022), R.S.~Lous (2018), M.~Jag (2016), C.~Kohstall (2012), A.~Trenkwalder (2011), S.~Spiegelhalder (2010), and E.~Wille (2009), available at \url{www.ultracold.at/grimm/theses/}. 

\subsubsection{Optically trapped mixture}
\label{sssec:ODT}

\begin{figure}
\centering
\vspace{-2mm}
\includegraphics[width=1.0\linewidth]{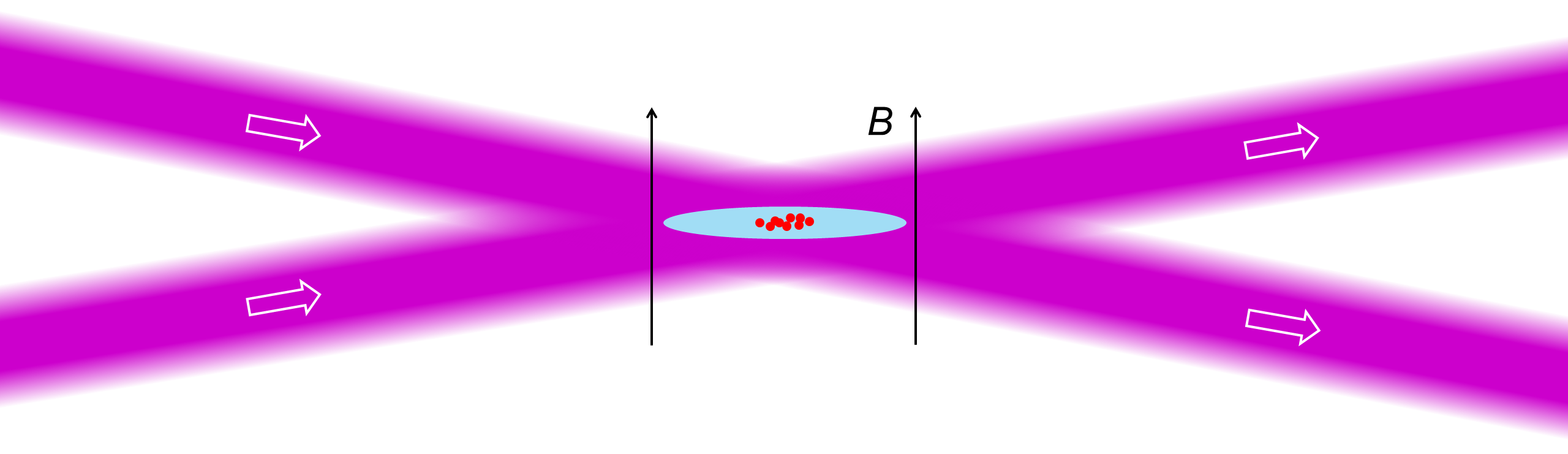}
\caption{Schematic experimental configuration. A crossed-beam optical dipole trap, realized with two near-infrared laser beams (wavelength 1064\,nm), contains a large degenerate cloud of $^6$Li fermions, in which K atoms are immersed as impurities. A magnetic field is applied to tune the interspecies interaction by means of a Feshbach resonance.}
\label{fig:LiKtrap}
\end{figure}

The mixture is prepared in a crossed-beam optical dipole trap \cite{Grimm2000odt} realized with near-infrared laser light. The trap, illustrated in Fig.~\ref{fig:LiKtrap}, contains typically $N_{\rm Li} \approx 10^5$ fermionic $^6$Li atoms in the lowest hyperfine and Zeeman sub-state at a temperature of the order of $T \approx 0.15 \, T_F$, where $T_F = \sqrt[3]{6 N_{\rm Li}} \, \hbar \bar{\omega}_{\rm Li}/k_B$ denotes the Fermi temperature defined for a harmonic trap. In our case, the mean trap frequency typically amounts to $\bar{\omega}_{\rm Li} / 2 \pi \approx 250$\,Hz, which results in a Fermi temperature of $T_F \approx 1\,\mu$K. The cigar-shaped Fermi sea in the trap (aspect ratio $\sim$$7.5$) has an axial (radial) size characterized by Thomas-Fermi radii of $\sim$$140\,\mu$m ($\sim$$18\,\mu$m).

The K impurities form a small cloud in the center of the trap, with a typical size much smaller than the large Fermi sea. This favorable fact is due to a combination of two different reasons: The Fermi pressure of the light $^6$Li atoms keeps the degenerate sea large, and the optical confinement of the impurity atoms is by a factor of $2.2$ stronger \cite{Lous2017toa}, owing to the ratio of dynamical polarizabilities at the trap wavelength. As a consequence, the impurity atoms ``see'' a nearly homogeneous environment, and effects of inhomogeneity stay negligibly small in all our experiments.

In the case of non-degenerate bosonic impurities (FB mixture), the situation is rather similar to the case of fermionic impurities (FF mixture). At sufficiently low temperatures, however, a Bose-Einstein condensate emerges \cite{Lous2017toa}. The BEC occupies a volume about a thousand times smaller than the Fermi sea. In this situation, the BEC can be considered as a mesoscopic impurity in a fermionic reservoir, as we will discuss later in Sec.~\ref{ssec:meso}.

\subsubsection{Impurity spin states}
\label{sssec:spinstates}

\begin{figure}
\centering
\vspace{0mm}
\includegraphics[width=0.9\linewidth]{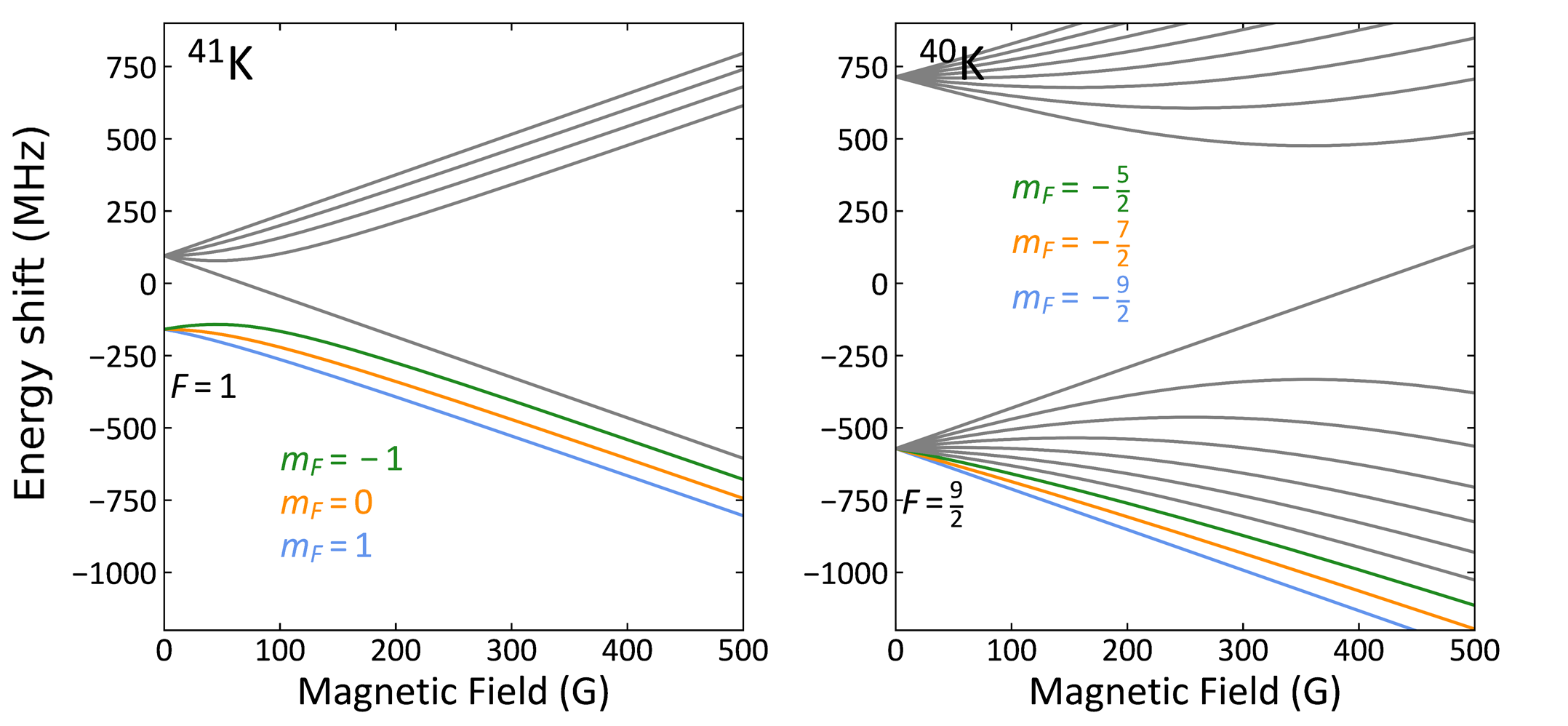}
\vspace{1mm}
\caption{Zeeman effect of the electronic ground states of $^{41}$K and $^{40}$K. In both cases, the three energetically lowest hyperfine spin states are relevant for our experiments.}
\label{fig:HF_splitting}
\end{figure}

While, in all our experiments, the Fermi sea is fully spin polarized with all $^6$Li atoms in the lowest internal state \footnote{In the preparation process, evaporative cooling is carried out with two spin components of $^6$Li, but one component is removed before the experiments are performed.}, the particular spin states of the K impurities and their manipulation by radio-frequency (RF) pulses play a crucial role.
The Zeeman effect of the ground states of $^{41}$K and $^{40}$K is illustrated in Fig.~\ref{fig:HF_splitting}. In both the BF and the FF mixture case, the lowest three sub-states are relevant for our experiments, and we introduce a unified notation for them according to their different roles in the experiments, see Tab.~\ref{tab:spin_states}.
While K$\ket{0}$ labels a state that features only the weak background interaction with the $^6$Li atoms, K$\ket{1}$ labels the state that offers tunability of the interspecies interaction 
via a magnetically controlled Feshbach resonance.
A third spin state, labeled K$\ket{\rm anc}$, is used as ancillary state in the preparation of the sample. It is used as a reservoir for impurity atoms, from which the state K$\ket{0}$ is filled in a controlled way (K$\ket{\rm anc}$$\rightarrow$K$\ket{0}$) before the actual experiments are carried out on the transition K$\ket{0}$$\rightarrow$K$\ket{1}$.

\begin{table}[t]
\caption{Unified notation for the relevant hyperfine spin states of $^{41}$K and $^{40}$K in terms of K$\ket{1}$, K$\ket{0}$, and K$\ket{\rm anc}$ (see text).
The alternative notation ``HF state'' labels the states in increasing energetical order. The quantum numbers $F$ and $m_F$ represent the common notation of hyperfine Zeeman sub-states.}
\label{tab:spin_states}
\vspace{-3mm}
\begin{center}
\resizebox{1.0\textwidth}{!}{
\begin{tabular}{c c c c|c c c c}
\hline
 notation & HF state& $F$ &  $m_F$ & notation & HF state& $F$ &  $m_F$ \\
\hline
K$\ket{1}$ & $^{41}\text{K}\ket{1}$&1 & 1 & K$\ket{\rm anc}$ & $^{40}\text{K}\ket{1}$ &9/2 & $-$9/2   \\
K$\ket{0}$ & $^{41}\text{K}\ket{2}$ &1 & 0 & K$\ket{0}$ &$^{40}\text{K}\ket{2}$&9/2 & $-$7/2  \\
K$\ket{\rm anc}$ & $^{41}\text{K}\ket{3}$ &1 & $-$1 & K$\ket{1}$ & $^{40}\text{K}\ket{3}$ & 9/2 & $-$5/2 \\
\hline
\end{tabular}
}
\end{center}
\end{table}

\subsubsection{Interaction tuning}
\label{sssec:tune}

\begin{figure}
\centering
\includegraphics[width=0.9\linewidth]{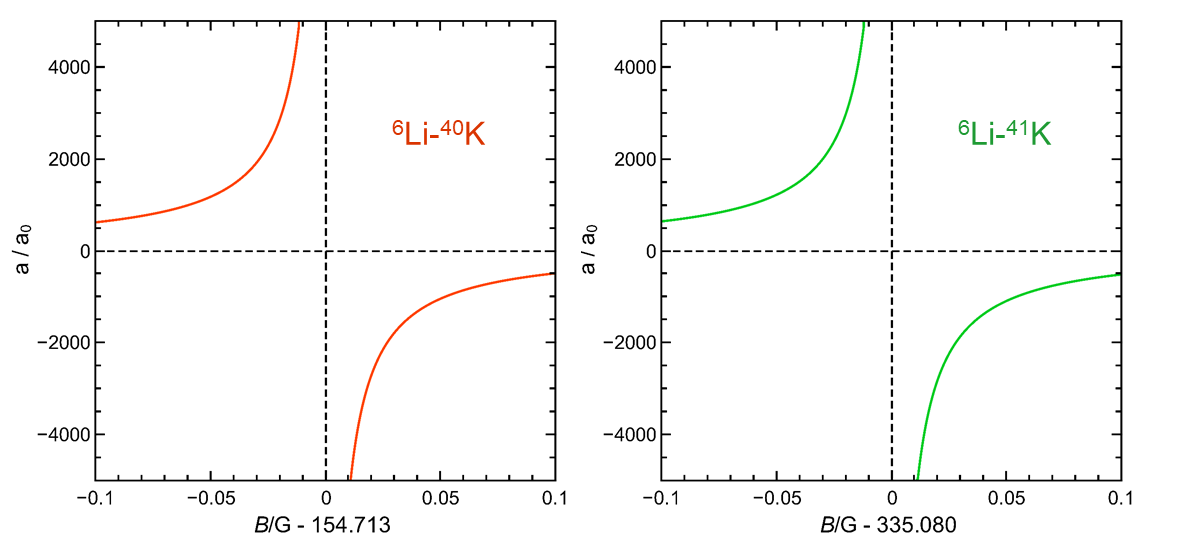}
\vspace{0mm}
\caption{Comparison of the two Feshbach resonances used in the experiments on FF and FB mixtures. Apart from being centered at different magnetic fields, their tuning behavior is very similar.}
\label{fig:resocompare}
\end{figure}

Feshbach resonances in the FF mixture of $^6$Li and $^{40}$K have been investigated in joint experimental and theoretical efforts in Refs.~\cite{Wille2008eau,Naik2011fri}. For the FB mixture of $^6$Li and $^{41}$K, experimental Feshbach scans can be found in Ref.~\cite{Lous2018PhD}, while theoretical calculations have been reported to us by T.~Hanna and E.~Tiesinga \cite{HannaPrivate} and by E.~Tiemann \cite{Tiemannpriv}. 

Two distinct resonances are of particular interest for our experiments. In the FF mixture, a resonance is found near $B_0 = 154.7$\,G in the spin channel that combines the lowest sub-state of $^6$Li with the third-to-lowest state of $^{40}$K. The corresponding parameter values have been determined 
\footnote{Values without specified uncertainties result from coupled-channel calculations based on experimental data obtained by Feshbach spectroscopy scans. Values with error bars result from additional measurements \cite{Cetina2015doi, Lous2018pti} of binding energies using radio-frequency techniques.}
to $\Delta = 0.88\,$G \cite{Naik2011fri}, $a_{\rm bg}=63.0\,a_0$ \cite{Naik2011fri}, and $\delta \mu
/ h = 2.66(2)\,$MHz/G \cite{Cetina2015doi}.
In the FB mixture, a resonance is found for both species in their lowest spin states near $B_0 = 335.1\,$G with parameter values \cite{Lous2018pti, Tiemannpriv} of $\Delta = 0.9487\,$G, $a_{\rm bg}=60.865\,a_0$, and $\delta \mu / h = 2.660(8)\,$MHz/G.

In Fig.~\ref{fig:resocompare}, we show the tuning behavior of the scattering length near both resonances, which hardly shows any difference. Also in terms of the range parameter values ($R^* = 2405(63)\,a_0$ for the FF case, and $R^* = 2241(7)\,a_0$ for the FB case) both resonances behave in a very similar way. It is a remarkable gift of nature that our system provides us experimentalists with two interspecies resonances with essentially the same two-body interaction properties. The Li-K mixture thus facilitates a direct comparison between fermionic and bosonic impurities interacting with the Fermi sea, as we will discuss in Sec.~\ref{sssec:mediated}.

It has been found experimentally \cite{Jag2014ooa, Lous2018pti} that the exact positions of the Li-K Feshbach resonances are subject to small shifts induced by the trap light itself. This light shift typically amounts to a few 10\,mG, which can significantly change the interaction in the resonance regime. Experimentally, this is a complication, but the effect can be tamed and used for very fast interaction tuning, as we have exploited in Refs.~\cite{Cetina2015doi, Cetina2016umb}, see also \ref{sssec:fastdyn}. 
 
We finally note that the intraspecies scattering lengths between different K atoms in various combinations of spin states are all similar and relatively small \cite{Lysebo2010fra, Ludewig2012fri, Regal2003mop, Loftus2002rco}, so that corresponding interaction effects remain weak and can be neglected in most cases of interest with resonant interspecies interaction.

\subsection{Fermi polarons}
\label{ssec:polarons}

The interest in polaron physics originated from pioneering experiments on strongly interacting fermionic spin mixtures with population imbalance \cite{Zwierlein2006doo, Partridge2006pap, Nascimbene2009coo}. The driving research question concerned the nature of the strongly interacting many-body system under conditions where fermionic pairing is inhibited by the mismatch of Fermi energies. The answer was found in the polaronic phase, where the minority atoms form quasiparticles dressed by excitations in the medium of the majority atoms. The physics of such polarons turned out to be remarkably rich. After more than a decade of research, it continues to produce new insights and surprises, and there is much more to come with novel mixed-species systems becoming available for experiments.

For an introduction to the theoretical description of Fermi polarons, we refer the reader to the lecture notes of M.\,M.~Parish and J.~Levinsen in these proceedings \cite{Parish2022fpa} and to various review articles \cite{Massignan2014pdm, Schmidt2018umb, Scazza2022rfa}.

In this Section, we review the key experiments on polaron physics that we carried out on Li-K mixtures. With this system we benefit from the large mass imbalance and from the possibility to change the quantum statistics of the impurity atoms without affecting other physical properties. Starting with the observation of the repulsive polaron (Sec.~\ref{sssec:repulsive}), we continue with the demonstration of fast impurity dynamics (Sec.~\ref{sssec:fastdyn}). We then turn our attention to the quantum statistics of impurities (Sec.~\ref{sssec:bosonic}). We finally present our very recent observations on mediated interactions between polarons (Sec.~\ref{sssec:mediated}), revealing striking differences between bosonic and fermionic impurity atoms.


\subsubsection{Observation of the repulsive polaron}
\label{sssec:repulsive}

Early experimental work \cite{Nascimbene2009coo, Schirotzek2009oof} focused on the Fermi polaron as the ground state of the system. However, on the repulsive side of a Feshbach resonance, the ground state is a weakly bound molecule formed by an impurity atom together with an atom from the fermionic medium. In this regime, the polaron can only exist as a \textit{metastable} state, and the key question was whether such a state would be sufficiently long-lived to be considered a well-defined quasi-particle in a meaningful sense for the description of repulsive many-body systems \cite{Massignan2014pdm}. Here, we summarize the main findings of Ref.~\cite{Kohstall2012mac} on our observation of the repulsive polaron. For a general review on repulsive polarons, see Ref.~\cite{Scazza2022rfa}.

\begin{figure}
\centering
\includegraphics[width=0.999\linewidth]{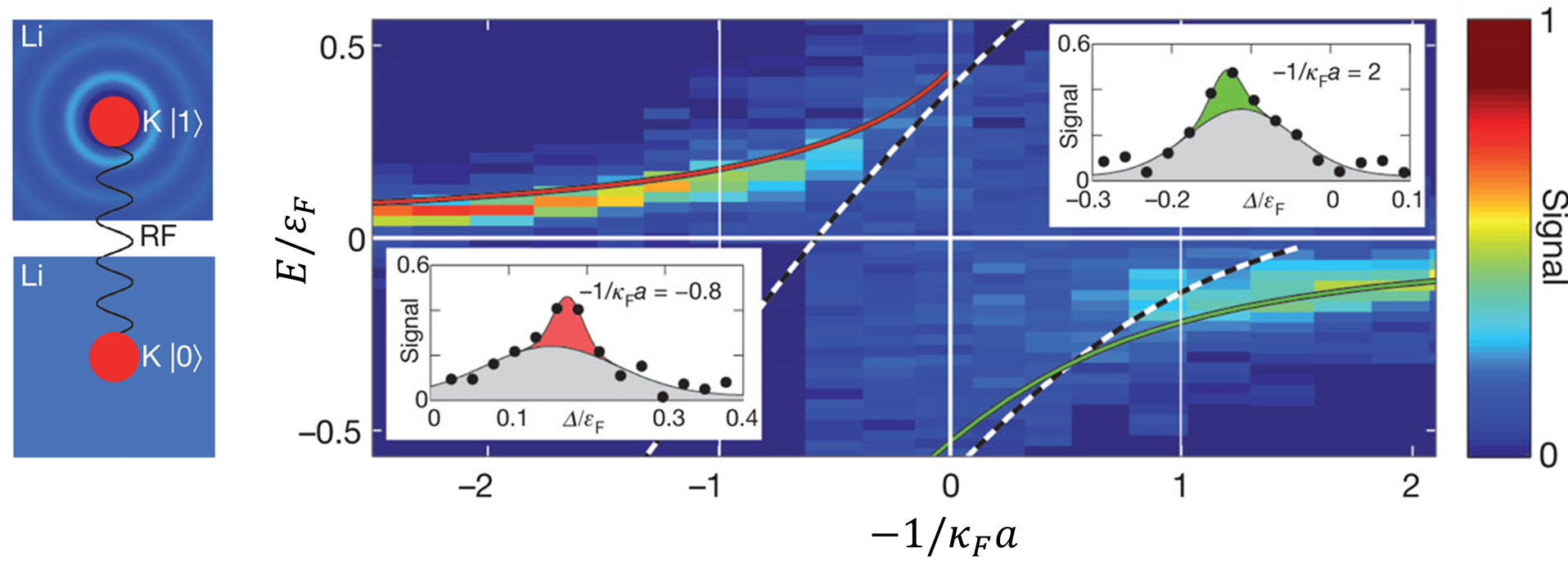}
\vspace{-3mm}
\caption{Observed polaron spectrum across a Feshbach resonance in the $^6$Li-$^{40}$K Fermi-Fermi mixture. The basic idea of the applied RF injection spectroscopy, transfering the impurity from a non-interacting to the interacting state, is illustrated on the left-hand side. The main panel shows a false-color plot of the spectrum across the resonance. Two selected spectra for the repulsive and the attractive case are represented by the insets. The solid lines refer to theory predictions of the polaron energies, and the area between the dashed lines corresponds to the molecule-hole continuum. Adapted from Ref.~\cite{Kohstall2012mac}.}
\label{fig:repulsive1}
\end{figure}

We used radio-frequency spectroscopic techniques \cite{Parish2022fpa, Vale2021spo} to probe the elementary properties of the metastable repulsive polaron. For recording the energy spectrum of the impurities, we employed ``injection'' spectroscopy (illustrated on the left-hand side of Fig.~\ref{fig:repulsive1}) by transferring $^{40}$K impurity atoms from the non-interacting state K$\ket{0}$ to the target state K$\ket{1}$ (Sec.~\ref{sssec:spinstates}). Our spectroscopic signal is the fraction of atoms transferred by an RF pulse 
\footnote{To optimize the signal, the power of the 1-ms RF pulses is chosen such that $\pi$-pulses are obtained for the bare K$\ket{0}$--K$\ket{1}$ transition.}
as a function of its detuning from the bare atomic resonance, recorded for different magnetic field strengths across the resonance.

To account for the universal character of the polaron problem, we express energies and interaction strengths in units determined by the Fermi sea of $^6$Li atoms. While the natural energy unit is the Fermi energy $\epsilon_F$ \footnote{To account for the residual inhomogeneity of the local Fermi energy $E_F({\bf r})$ experienced by the impurity atoms, we define an effective Fermi energy $\epsilon_F$ as an average over the impurity cloud.}, we introduce the dimensionless interaction strength parameter $-1/\kappa_F a$, where $\kappa_F =\sqrt{2 m_{\rm Li} \epsilon_F} / \hbar$ is the Fermi wave number and $a$ is the tunable $s$-wave scattering length. According to our definition, negative (positive) values of the interaction parameter correspond to repulsive (attractive) interspecies interactions.
The particular Feshbach resonance (see discussion in Sec.~\ref{sssec:tune}) is further characterized by the dimensionless range parameter $\kappa_F R^*$. In the experiments discussed here \cite{Kohstall2012mac}, the parameter values are $\epsilon_F = h\times 37$\,kHz, $\kappa_F = 1/(2850\,a_0)$, and $\kappa_F R^*= 0.95$.

Typical spectroscopic signals are shown in the insets of Fig.~\ref{fig:repulsive1} for repulsive and attractive interactions. The polaron manifests itself as a narrow peak on a broad pedestal, the latter resulting from excitations in the Fermi sea. The polaron energy is given by the interaction-induced frequency shift of the peak. The whole energy spectrum recorded across the resonance is shown as false-color plot in the main panel. The two polaron branches are clearly visible on the repulsive and the attractive side of the resonance. The signal gets weaker on both sides when the resonance is approached, which marks the gradual disappearance of the polaron peak. Close to resonance, where the interaction is very strong, the signal becomes very weak, and a detection of the spectrum  requires much stronger pulses \cite{Kohstall2012mac} or more advanced spectroscopic methods as described in Sec.~\ref{sssec:fastdyn}.

The essential features of the observed polaron spectrum agree well with predictions from theory \cite{Chevy2006upd, Punk2009ptm, Massignan2011rpa}, based on a variational approach that considers single particle-hole excitations in the Fermi sea (see also the contribution of M.~Parish and J.~Levinsen to these proceedings \cite{Parish2022fpa}). The solid lines show the theoretical behavior of the polaron energy versus interaction strength, matching the experimental observations. The coupling to molecular excitations happens in the molecule-hole continuum (MHC), which is delimited by the dashed lines. 
Here an impurity atom can form a molecule with any atom from the bottom to the top of the Fermi sea.
In the MHC no polaron peak is observed, but a rather broad spectrum of excitations.

\begin{figure}
    \centering
    \vspace{0mm}
        \includegraphics[width=1.00\textwidth]{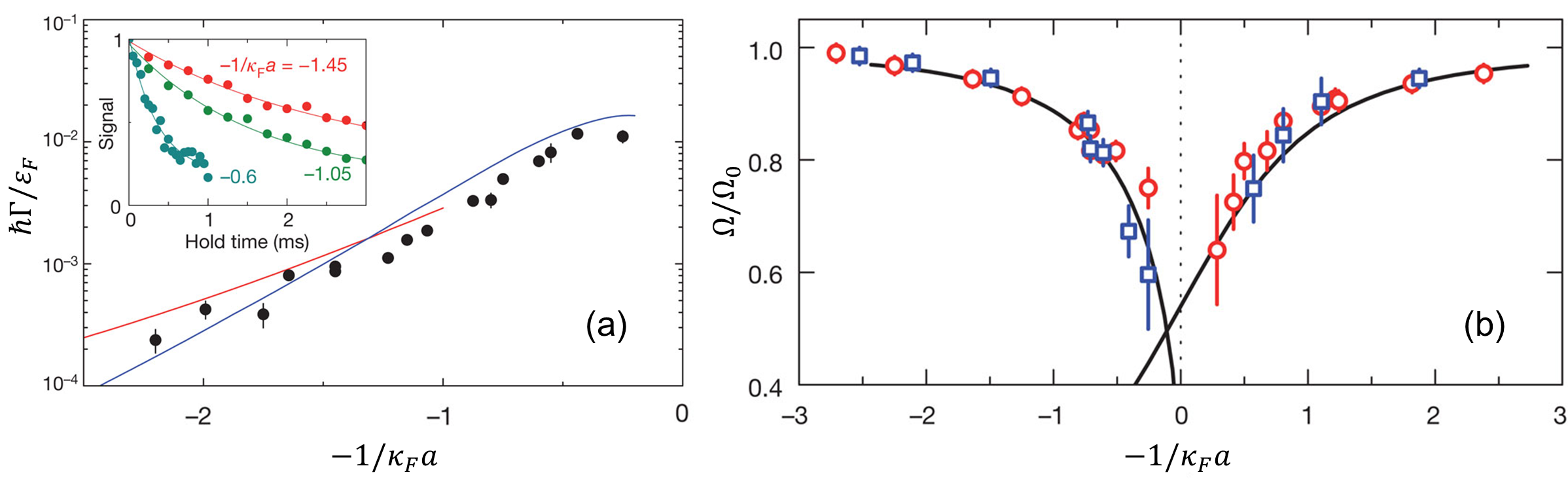}
    \vspace{-3mm}
    \caption{Key properties of the repulsive polaron. In (a), measured values for the decay rate $\Gamma$ (in units of the inverse Fermi time $\epsilon_F / \hbar$) are compared with the predictions from theoretical models for two-body decay (long solid blue line) and three-body decay (shorter red line). The inset shows decay curves for three different values of the interaction strength.
    In (b), measurements of the interaction-reduced Rabi frequency $\Omega$ normalized to the bare Rabi frequency $\Omega_0$ are shown in comparison with a theoretical calculation of $\sqrt{Z}$ (black solid lines), where $Z$ is the quasiparticle residue of the polaron. For comparison, we also show the corresponding behavior for the attractive polaron (positve values of the interaction parameter).
    Adapted from Ref.~\cite{Kohstall2012mac}.}
    \label{fig:repulsive2}
\end{figure}

The lifetime of the repulsive polaron is crucial for its existence as a well-defined quasiparticle. In the experiments, we studied the lifetime employing a two-pulse technique, in which, after creating the polaron with a first RF pulse, it was converted back to the bare particle in a second pulse. In between these two RF pulses, a variable hold time was introduced, during which decay into the MHC took place. This decay of the polaron into molecular excitations inhibited the back-conversion. Typical decay curves, recorded in this way, are shown in the inset of Fig.~\ref{fig:repulsive2}(a). The measured decay rates are shown in the main panel of (a) in comparison with the results of theoretical model calculations \cite{Massignan2011rpa, Bruun2010dop} for two-body decay (long solid blue line) and three-body decay (shorter solid red line). Our results demonstrated that the repulsive polaron can have a rather long lifetime, i.e.\ much longer than the Fermi time $\hbar/\epsilon_F$, and thus answered the question on the existence of this well-defined metastable quasiparticle positively. We note that, for the case of a repulsive FF mixture, the polaron lifetime is highly relevant in view of the emergence of itinerant ferromagnetism, which has been a debated topic \cite{Massignan2014pdm, Jo2009ifi, Valtolina2017etf, Amico2018tro}.

In Ref.~\cite{Kohstall2012mac}, we also introduced a new spectroscopic method based on the observation of Rabi oscillations, induced by a continuous RF drive. We showed that the \textit{quasiparticle residue} $Z$, which quantifies the bare-particle component in the polaron wavefunction \cite{Parish2022fpa}, can be derived from a reduction of the Rabi frequency acccording to 
$\Omega = \sqrt{Z} \Omega_0$, where $\Omega_0$ is the unperturbed Rabi frequency. This method, which has become an important tool for experiments on various systems \cite{Parish2022fpa, Scazza2017rfp, Adlong2020qlo}, provides an alternative way to measure $Z$ instead of determining the area under the polaron peak. The latter method requires linear-response conditions, which can only be achieved at very low power with weak signals \footnote{For a discussion of linear response measurements under realistic experimental conditions, see the Supplementary Material of Ref.~\cite{Cetina2016umb}.}.

\subsubsection{Ultrafast dynamics}
\label{sssec:fastdyn}

Out-of-equilibrium dynamics of fermionic many-body systems is at the heart of many problems in science and technology. The Fermi energy sets the shortest time scale of the problem, referred to as the Fermi time $\tau_{F} = \hbar/\epsilon_{\rm F}$. In a metal (Fermi sea of electrons), it is of the order of 100 attoseconds, making the experimental observation of \textit{ultrafast} (faster than the Fermi time) dynamics extremely challenging. 
Ultracold gases feature much lower Fermi energies, corresponding to Fermi times in the range of a few microseconds. This brings the ultrafast dynamics into an experimentally accessible range, where powerful spectroscopic techniques, such as atom interferometry \cite{Cronin2009oai}, can be applied.

\begin{figure}
    \centering 
    \includegraphics[width=0.75\textwidth]{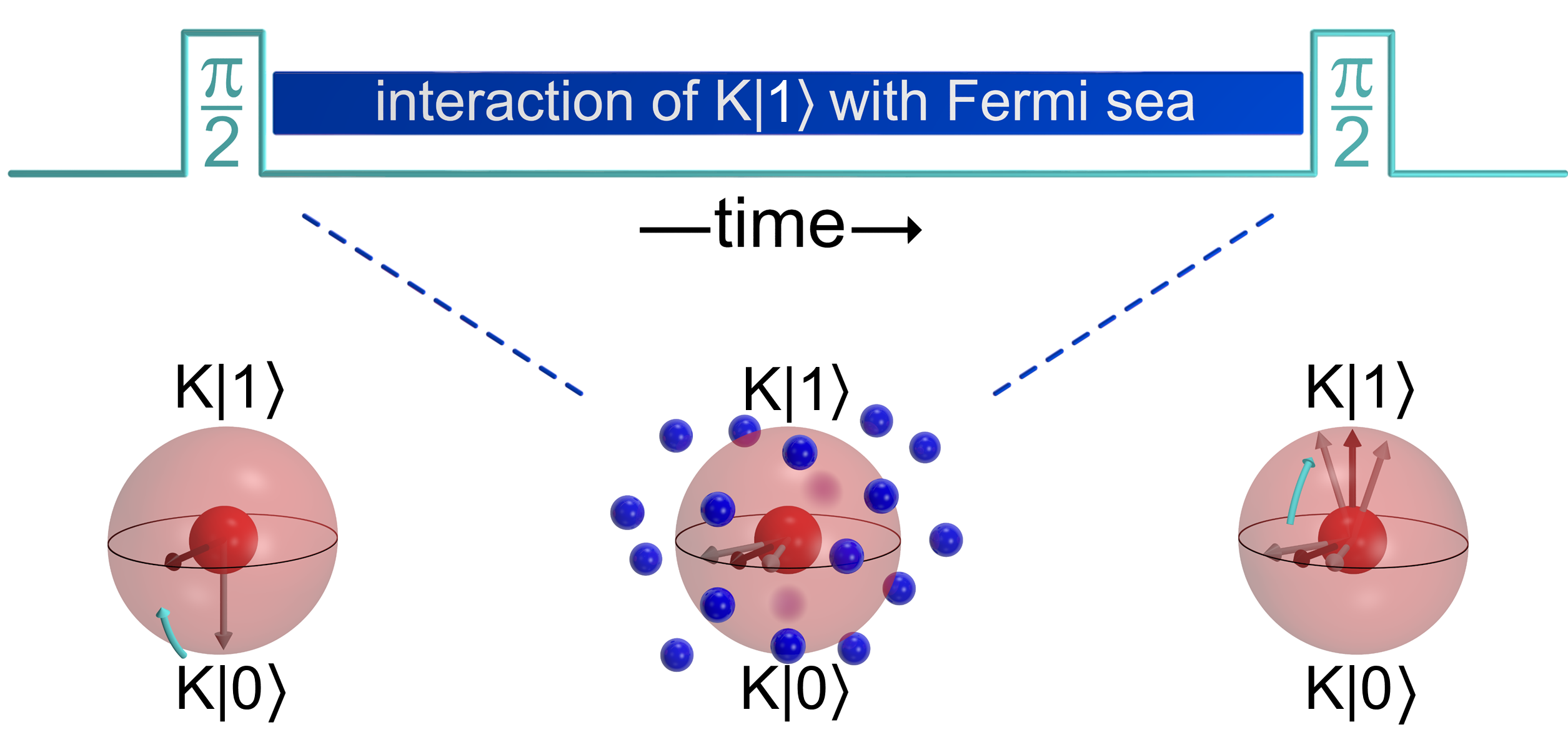}
    \caption{Illustration of Ramsey interferometry. The sequence starts with a first RF $\pi/2$ pulse, which is applied in the presence of weak interactions between the impurity atoms and the Fermi sea. As illustrated on the Bloch sphere, this pulse drives an impurity atom (big red dot) into a superposition of the spin states K$\ket{0}$ and K$\ket{1}$. By optical resonance shifting (see text), the interaction of the K$\ket{1}$ component with the atoms of the Fermi sea (small blue dots) is abruptly turned on while the K$\ket{0}$ component remains noninteracting. The impurity state then evolves for a variable interaction time, at the end of which its state is probed by a second $\pi/2$ pulse and subsequent measurement of the spin-state populations.
    Adapted from Ref.~\cite{Cetina2016umb}.}
    \label{fig:Ramsey}
\end{figure}

In Ref.~\cite{Cetina2016umb}, we studied the ultrafast dynamics of a system of $^{40}$K impurities immersed in the $^6$Li Fermi sea. We measured the response of the system to a suddenly introduced impurity \cite{Goold2011oca, Knap2012tdi}, using the spectroscopic method of Ramsey interferometry \cite{Vale2021spo, Knap2012tdi}. The basic idea of this powerful method is illustrated in Fig.~\ref{fig:Ramsey}. The impurity atom, being  initially in the noninteracting state K$\ket{0}$, is transferred by a first RF $\pi/2$ pulse into a coherent superposition of K$\ket{0}$ and K$\ket{1}$. It then evolves freely for a variable interaction time, before a second $\pi/2$ pulse is applied with a variable phase $\phi$. Finally, the population difference in the two spin states K$\ket{0}$ and K$\ket{1}$ is measured as a function of $\phi$, which yields our `Ramsey signal'. This signal yields the time-dependent complex-valued overlap function $S(t) = |S(t)| \exp[-i\varphi(t)]$, which is discussed in more detail in Refs.~\cite{Parish2022fpa, Knap2012tdi}.

The conditions of our Ramsey-type experiments were close to the injection spectroscopy measurements discussed before in Sec.~\ref{sssec:repulsive}, with a Fermi energy of $\epsilon_F = k_B \times 2.6\,\mu{\rm K} = h \times 54\,$kHz, corresponding to  a Fermi time of $\hbar/\epsilon_F = 2.9\,\mu$s, and a temperature of $T\approx 0.17\, T_F$. We had to face the problem that the application of short RF $\pi/2$~pulses is in practice limited by the available RF power to typically $10\,\mu$s, clearly longer than the Fermi time. In this case, the interaction with the Fermi sea during the pulse complicates the situation. To overcome this problem we applied a special trick, exploiting the optical shift of the Feshbach resonance that is induced by the trap light \cite{Cetina2015doi}, which is much faster than possible with magnetic-field changes. We switched the depth of the optical trap, keeping the trap frequency essentially constant. In this way, as illustrated in Fig.~\ref{fig:Ramsey}, we applied the RF pulses under weakly interacting conditions and tuned to strong interactions only between the two pulses.

\begin{figure}
    \centering 
    \includegraphics[width=0.999\textwidth]{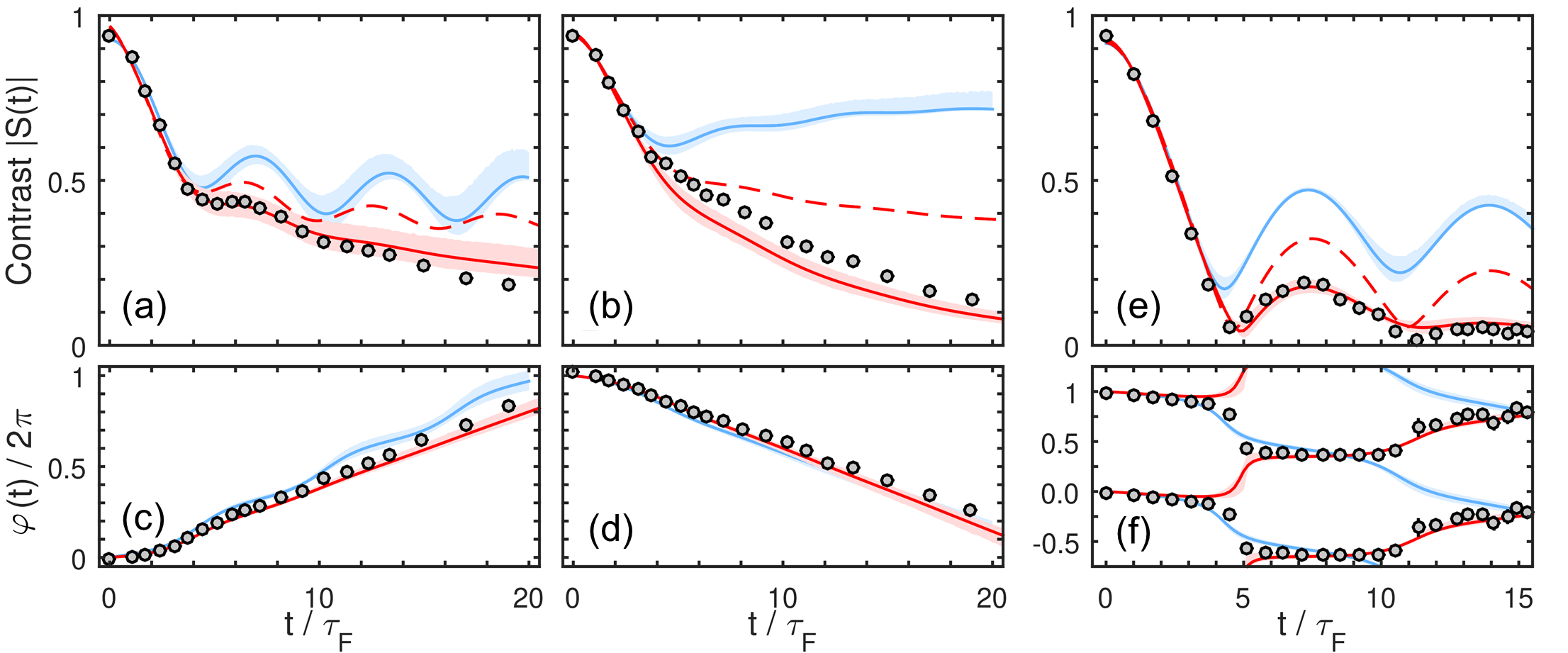}
    \caption{Impurity dynamics in the Fermi sea. Contrast and phase of the interference signal as a function of the reduced interaction time $t/\tau_{F}$ in (a, c) the repulsive polaron regime for $-1/\kappa_F a = -0.23$, in (b, d) in the attractive case for $-1/\kappa_F a = 0.86$, and (e, f) for resonant interactions ($-1/\kappa_F a = 0.08$).
    The lines refer to different theoretical approaches (see text).
    Adapted from Ref.~\cite{Cetina2016umb}.}
    \label{fig:ultrafast}
\end{figure}

In Fig.~\ref{fig:ultrafast}(a-d), we show the observed time evolution of $S(t)$ for the interaction conditions for which our earlier experiments (Sec.~\ref{sssec:repulsive}) had demonstrated the presence of repulsive and attractive polarons. In both cases, the contrast signals $|S(t)|$ show an initial parabolic transient within an evolution time of $\sim$$4\,\tau_F$. Polarons essentially emerge in this short time, in which the dressing cloud is formed by excitations deep in the Fermi sea. We thus interpret this result as \textit{real-time observation of polaron formation}. For longer times, an exponential decay of the contrast signal marks the thermal decoherence of the polaron, as we have studied in more detail using a spin-echo technique \cite{Cetina2015doi}.

On resonance, for the strongest possible interactions, a description of the dynamics in terms of a single dominant quasiparticle excitation breaks down. In this regime, our measurements displayed in Fig.~\ref{fig:ultrafast}(e, f) reveal the striking quantum dynamics of an interacting fermionic system forced into a state far out of equilibrium. The contrast $|S(t)|$ shows pronounced oscillations with minima reaching almost zero, which indicates
that the time-evolved state can become almost orthogonal to the initial state. Meanwhile,
the phase $\varphi(t)$ exhibits plateaus with jumps of $\pi$ near the contrast minima.
This behavior can be interpreted in terms of a quantum beat between the repulsive and the attractive branch of our many-body system. The two branches coexist in a coherent superposition, but they are strongly broadened, which results in strong damping of the oscillations.

Our experimental results served as an important benchmark for theoretical models.
The functional determinant approach \cite{Schmidt2018umb, Knap2012tdi} provides an exact solution for a fixed (immobile) impurity at arbitrary temperatures, taking into account the nonperturbative creation of infinitely many particle-hole pairs in the Fermi sea. 
In Fig.~\ref{fig:ultrafast}, a corresponding calculation is shown by the solid red lines. We see excellent agreement with our experimental results, which indicates that the
effects of impurity motion remain small in our system. This observation can be explained by
the fact that our impurity is sufficiently heavy so that the effects of its recoil with energies are masked by thermal fluctuations. To identify the effect of temperature, we
performed a corresponding calculation assuming $T = 0$; the results are shown as dashed red lines in Fig.~\ref{fig:ultrafast}. In the zero-temperature case, a slower decay of $|S(t)|$ would be expected, which follows a power law at long times under the idealizing assumption of infinitely heavy impurities.

An alternative description is based on a time-dependent version of the variational approach that was successfully applied to describe the spectral properties of the polaron, see Sec.~\ref{sssec:repulsive}. The model considers only single particle-hole excitations, neglecting higher-order effects in the thus truncated basis. Corresponding results for the  zero-temperature case, shown by the solid blue lines in Fig.~\ref{fig:ultrafast}, capture the main features at shorter interaction times, but deviate from the experiments at longer times. In Ref.~\cite{Liu2019vaf} the variational approach was generalized to finite temperatures, and the results (see also Ref.~\cite{Parish2022fpa} in these proceedings) show excellent agreement with the experiments.

The two discussed models rely on different assumptions, but benchmarked against the present experimental results they appear to perform equally well. We neither see deviations connected with the impurity mobility (in comparison with the functional determinant approach) nor effects of multiple particle-hole excitations (in comparison with the variational approach). Because of their low energy scales, both effects are expected to modify the signal only at longer interaction times. Here, however, the behavior is dominated by the finite temperature of the system, which masks all other effects.

A challenging prospect of future work with time-domain spectroscopy is the observation of Anderson's orthogonality catastrophe \cite{Knap2012tdi, Anderson1967ici}. As a paradigm in condensed-matter physics, it refers to the fact that even a small local perturbation of a Fermi sea (as e.g.\ induced by an impurity) can lead to a strong reduction of the wavefunction overlap between many-body states. Experimentally, this would require immobile impurities, which can be achieved by pinning them in a species-selective optically lattice \cite{LeBlanc2007sso, Lamporesi2010sim}, along with lower temperatures. In Ramsey-type interferometry experiments, the signature would be a power-law decay instead of the exponential decay \cite{Knap2012tdi}, but also other techniques \cite{Adlong2021sot} have been proposed to reach this challenging goal.

\subsubsection{Bosonic impurities}
\label{sssec:bosonic}

Potassium offers the remarkable possibility to choose the impurity quantum statistics by working with the fermionic isotope $^{40}$K or the bosonic isotope $^{41}$K~\footnote{The other stable bosonic isotope of potassium, $^{39}$K, features different laser cooling and ultracold interaction properties, which make it less suited for the present experiments.}. The fact that, in both cases, Feshbach resonances of very similar character (see Sec.~\ref{sssec:tune}) are available for tuning the interaction with the Fermi sea of $^6$Li is a remarkable gift of nature, which facilitates a direct comparison of the FF with the FB system. Our first experiments on polarons formed by bosonic impurities in a fermionic medium are reported in Ref.~\cite{Fritsche2021sab}. Here we give a short summary of the main findings, which also paved the way for the more refined experiments reported in Sec.~\ref{sssec:mediated}.

\begin{figure}
    \centering 
    \includegraphics[width=\textwidth]{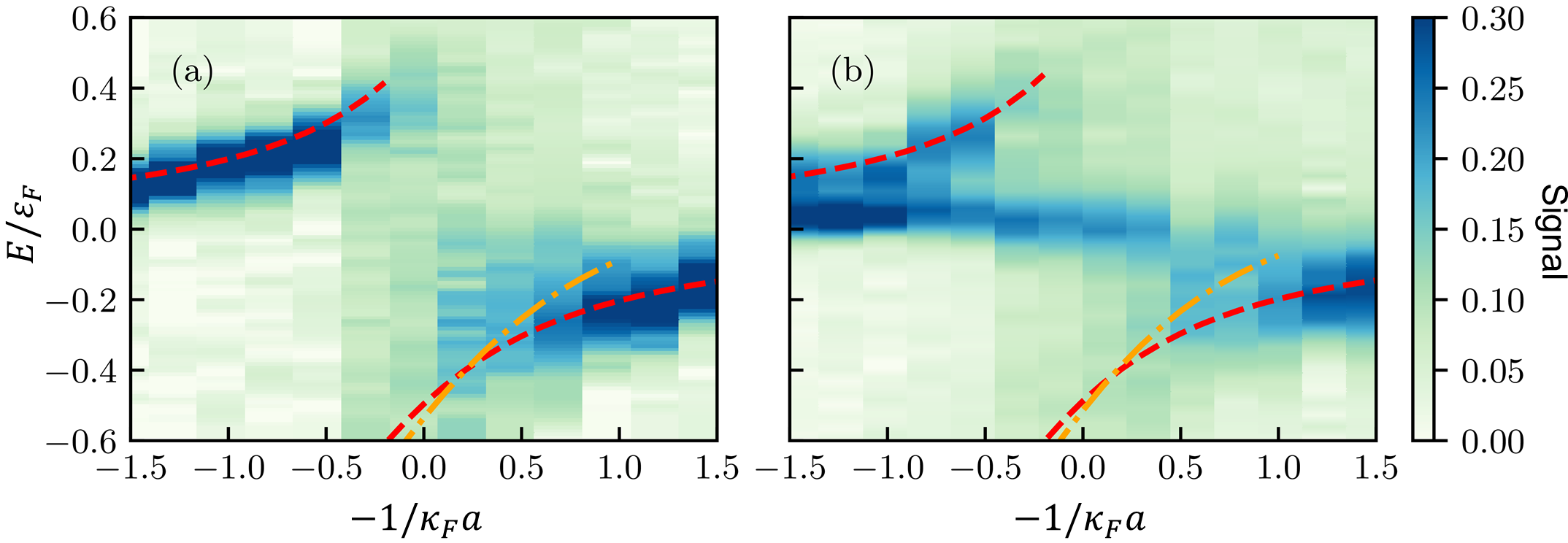}   
    \caption{Spectral response of bosonic $^{41}$K impurities immersed in a $^6$Li Fermi sea.
    Panels (a) and (b) show the measured excitation spectra for $T/T_F \approx 0.14$ (thermal cloud regime)  and $T/T_F \approx 0.19$ (partially condensed regime), respectively. The energy spectra (normalized energy $E/\epsilon_F$) are shown as a function of the dimensionless interaction parameter. The signal obtained by injection spectroscopy refers to the fraction of atoms transferred from the non-interacting state K$\ket{0}$ to the Feshbach-resonant state K$\ket{1}$. The dashed red and dash-dotted orange lines illustrate theoretical predictions for the polaron and molecule energies in the single-impurity limit, respectively. Adapted from Ref.~\cite{Fritsche2021sab}.}
    \label{fig:spectraFB}
\end{figure}

In the single-impurity limit, the impurity quantum statistics is irrelevant and the FF and FB systems are expected to behave in the same way. Accordingly, their spectral functions should be one and the same in the dilute limit of low impurity concentration. This is confirmed by our RF injection spectroscopy experiments, as one can see by comparing  Fig.~\ref{fig:spectraFB}(a) with Fig.~\ref{fig:repulsive1}, the latter discussed in Sec.~\ref{ssec:polarons}. We could also confirm that the lifetime behavior of the metastable repulsive polaron is essentially the same for fermionic and bosonic impurities \cite{Kohstall2012mac, Fritsche2021sab}.

However, the bosonic $^{41}$K impurity cloud can reach the condition for Bose-Einstein condensation, as we discuss in more detail later in Sec.~\ref{ssec:meso}. While the spectrum shown in Fig.~\ref{fig:spectraFB}(a) was recorded for thermal impurities close to degeneracy, the spectrum in (b) was taken for a slightly colder sample, where a partial BEC was formed. As can be seen clearly, polaronic branches still exist, but an additional branch shows up, which connects both sides with very small interaction shifts. These observations can be readily interpreted considering the inhomogeneous situation in a harmonic trap. The thermal fraction of impurities forms a shell around the BEC located in the very center of the trap. While the thermal part behaves like the polaronic situation in (a), the interaction physics is completely different for the BEC. Here the local number density of the K atoms is much larger (about 30 times) than the one of the Li atoms. Consequently, impurities and bath interchange their roles, and we observe a K-BEC doped with Li impurities. This situation can be interpreted in terms of Bose polarons \cite{Hu2016bpi, Jorgensen2016ooa, Penaardila2019aab, Yan2020bpn, Skou2021neq}, which is also supported by an estimate of the interaction energy \cite{Fritsche2021sab}.
In our FB mixture, we thus observed that both Fermi and Bose polarons can exist in different regions of the same trap.

\subsubsection{Mediated interactions}
\label{sssec:mediated}

Beyond the single-impurity limit, quasiparticles can interact with each other via modulation of the surrounding medium. Conventional superconductivity, for example, arises from phonon-mediated interactions between electrons. The interaction between quasiparticles due to the exchange of particle-hole excitations in a surrounding Fermi sea is crucial for understanding both the equilibrium and dynamical properties of Fermi liquids. It is of particular importance for the emergence of collective modes in Fermi liquids \cite{Baym2004book}, the appearance of giant magneto-resistance, as well as for the coupling between nuclear magnetic moments as predicted by Ruderman-Kittel-Kasuya-Yosida (known as RKKY interaction) \cite{Nozieres1964book}.

The experimental observation of mediated interactions between ultracold atoms in a Fermi sea is in general very challenging. For a BEC of Cs atoms immersed in a Fermi sea of $^6$Li, this was recently achieved \cite{DeSalvo2019oof} under conditions of weak interspecies interaction, i.e.\ not in the regime where the formation of quasiparticles becomes relevant. In the strongly interacting regime, our experiments showed some hints on mediated interactions between quasiparticles \cite{Cetina2016umb, Fritsche2021sab}, but an unambigous detection of such effects had to wait until we carried out a new generation of experiments under substantially improved conditions \cite{Baroni2023mib}, after heaving reduced various sources of fluctuations in the experimental set-up. Here we summarize the main findings of Ref.~\cite{Baroni2023mib} on the observation of the effect of mediated interactions for both fermionic and bosonic impurity atoms.

We define the impurity concentration ${\mathcal C}=n_\downarrow/n_\uparrow$, where $n_\downarrow$ and $n_\uparrow$ represent the number densities of the impurity atoms and the medium.
According to Landau's Fermi-liquid theory \cite{Yu2012iii}, for small $\mathcal{C}$, we can write the energy needed to create a polaron as \cite{Fritsche2021sab, Baroni2023mib}
\begin{equation}
\epsilon_{\downarrow}=\epsilon_{\downarrow}^0 + s\,{\mathcal C} \epsilon_F \, ,
\label{eq:QPenergy}
\end{equation} 
where $\epsilon_{\downarrow}^0$ represents the energy of a polaron in the single-impurity limit. To characterize the universal polaron-polaron interaction, we have introduced the dimensionless quantity $s$ to which we refer as \textit{mediated interaction coefficient}. It describes the \textit{linear} dependence on  the impurity concentration.
In the limit of vanishing energy and momenta, this coefficient can be expressed in a remarkably simple form as
\begin{equation}
s = \mp \frac{2}{3}(\Delta N)^2 \, ,
\label{eq:scoeff}
\end{equation}
where $\Delta N$ represents the number of particles in the dressing cloud around a zero-momentum impurity \cite{Massignan2011rpa}, and the minus (plus) sign refers to bosonic (fermionic) impurities.
Accordingly, the sign of the coefficient $s$ reveals a very interesting behavior: 
\begin{itemize}
\item Because of the quadratic dependence on $\Delta N$, it does not matter whether the interaction with the medium is repulsive ($\Delta N <0$) or attractive  ($\Delta N >0$). This means that repulsive  or attractive polarons show basically the same mediated interaction behavior. 
\item The sign of $s$ is determined by the impurity quantum statistics. While Fermi polarons made of bosonic impurities always attract each other, polarons made of fermionic impurities always repel each other.
\end{itemize}

The mediated interaction effect stays rather weak, since $|\Delta N| \lesssim 1$ \cite{Massignan2011rpa}. Therefore the interaction-induced energy shift relative to the Fermi energy, as quantified by $s \mathcal{C}$, amounts to typically just a few percent. This makes the experimental observation highly challenging. 

To measure the small expected shifts, we applied RF injection spectroscopy in basically the same way as described in Secs.~\ref{sssec:repulsive} and \ref{sssec:bosonic}, but with substantially improved stability of our set-up against various sources of fluctuations, mostly magnetic-field and atom number fluctuations. The fluctuations were monitored and data were rejected if they exceeded certain thresholds. We recorded a large set of spectra for variable values of the concentration $\mathcal{C}$ and the interaction parameter $-1/\kappa_F a$, for both bosonic impurities in the FB mixture and fermionic impurities in the FF mixture. The whole process of data acquisition took several months. 

The experimental conditions were similar to our previous experiments, with $\epsilon_F/h \approx 16\,$kHz (20\,kHz), $\kappa_F R^* \approx 0.54$ ($0.62$), and $T/T_F \approx 0.15$ ($0.25$) for the FB (FF) mixture. For more details of the experimental procedures and the data analysis, see Ref.~\cite{Baroni2023mib}.

\begin{figure}
\centering 
\includegraphics[width=0.80\textwidth]{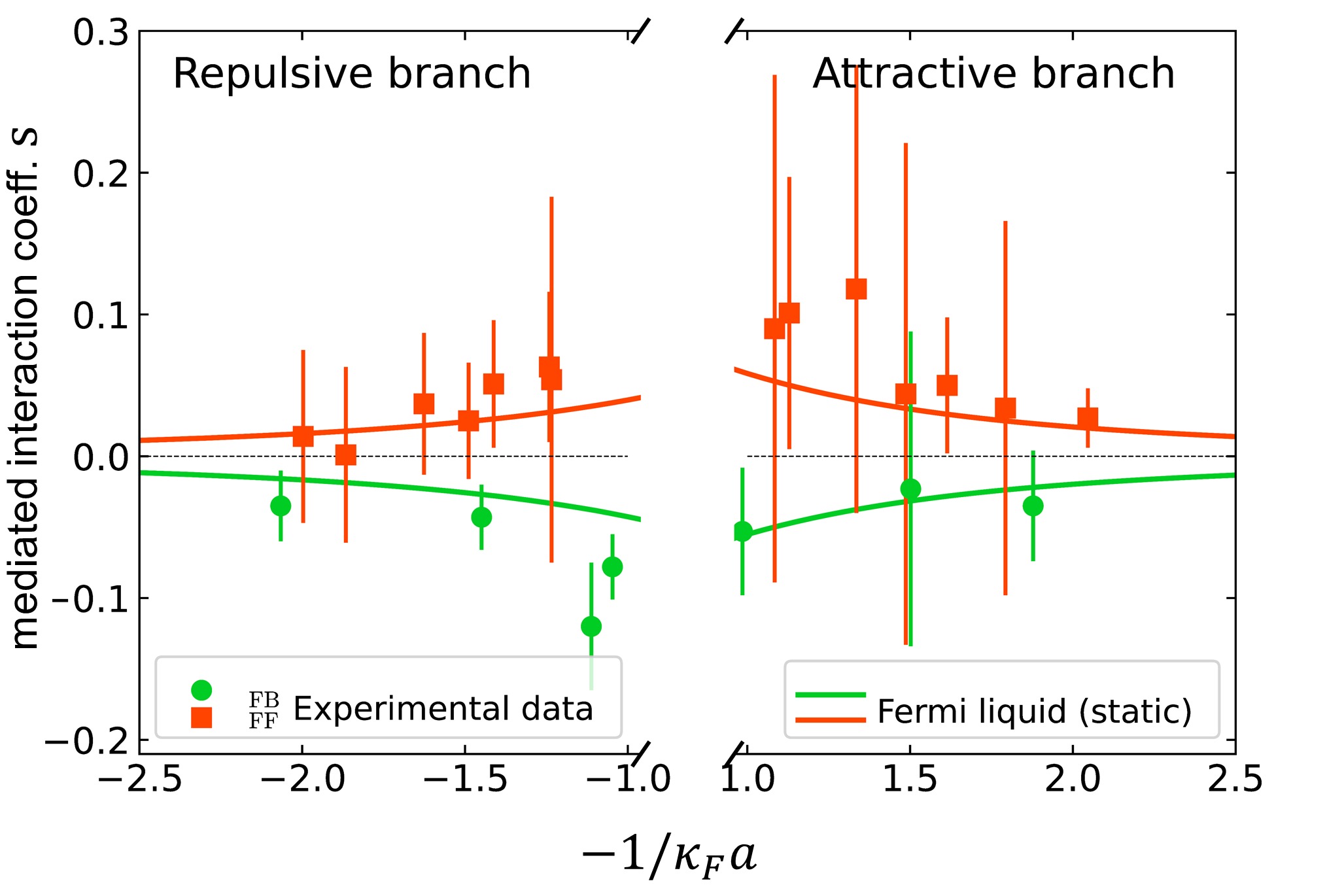}
\caption{Mediated interaction coefficient $s$ in the regime of repulsive and attractive polarons for moderate interaction strength ($ |1/k_Fa| \geq 1$). The filled red squares (filled green dots) refer to measurements on fermionic $^{40}$K (bosonic $^{41}$K) impurities. The solid lines show corresponding predictions according to Eq.~(\ref{eq:scoeff}). The measurements verify the curious behavior of the sign, which depends on the impurity quantum statistics, but not on the attractive or repulsive character of the polaron.
Adapted from Ref.~\cite{Baroni2023mib}.}
\label{fig:mediated}
\end{figure}

In Fig.~\ref{fig:mediated}, we show our main results of measurements taken for $|1/k_Fa| \geq 1$. In this regime of moderate interactions, i.e.\ close, but not too close to resonance, our previous results have confirmed the polaronic character with a large quasiparticle residue $Z \gtrsim 0.8$, see Fig.~\ref{fig:repulsive2}(b). The values measured for the interaction coefficient $s$ indeed exhibit the expected behavior of the sign, being the same for the repulsive and the attractive side of the resonance, but changing with the impurity quantum statistics. Our results confirm the key prediction of Fermi-liquid theory that polarons formed with bosonic (fermionic) impurities attract (repel) each other. Also quantitatively, the observed behavior is consistent with the prediction of Eq.~(\ref{eq:scoeff}).

We have carried out additional measurements closer to resonance  (not shown), as discussed in detail in Ref.~\cite{Baroni2023mib}. In the strongly interacting regime ($|1/\kappa_Fa|<1$), we observed striking deviations from the behavior described by Eq.~(\ref{eq:scoeff}). This includes a sign reversal on the attractive side, which suggests an effect caused by the formation of dressed molecules \cite{Punk2009ptm,Trefzger2012iia, Massignan2012pad}. We also observed a breakdown for repulsive interactions when the polaron becomes ill-defined with decreasing residue. We found that an extended many-body theory, based on the ladder approximation generalized to non-zero impurity concentrations \cite{Baroni2023mib}, can partially explain these observations, but more theoretical and experimental studies are required to fully understand the intricate interaction physics in the regime of very strong interactions, which may involve intriguing physics well beyond the Fermi-liquid 
paradigm. 

Moreover, at concentrations $\mathcal{C} \approx 1$, it would be interesting to explore the general connection of the polaron picture with nearly balanced FB \cite{Fratini2010pac, Ludwig2011qpt, Yu2011sco, Duda2023tfa} and FF \cite{Nascimbene2011flb, Gubbels2013ifg, Pini2021bmf} mixtures.

\subsection{BEC as a mesoscopic impurity}
\label{ssec:meso}

In the preceding Sections \ref{sssec:bosonic} and \ref{sssec:mediated}, we have seen that bosonic impurity atoms in a Fermi sea behave quite  similar to fermionic ones, as long as they stay in the thermal regime, and the only notable difference is the sign of mediated interactions. This changes dramatically if the temperature is low enough for Bose-Einstein condensation. The small-sized BEC can then be considered as a single mesoscopic impurity with a number density that largely exceeds the number density in the Fermi sea. As an interesting consequence, we have already discussed the appearance of a new branch in the RF spectrum, see Fig.~\ref{fig:spectraFB}(b). But there are other striking consequences if the interaction is tuned into the strongly repulsive regime. The BEC phase separates from the Fermi sea with dramatic effects in its collective behavior.

Here we first discuss the application of a BEC as an accurate and very sensitive thermometer for the Fermi gas at the lowest achievable temperatures (Sec.~\ref{sssec:BECtemp}). We then turn our attention to the phase separation effect and its manifestations in the static and dynamic behavior for strongly repulsive interactions (Secs.~\ref{sssec:phasesep} and \ref{sssec:breathing}).

\subsubsection{BEC thermometry}
\label{sssec:BECtemp}

In experiments on deeply degenerate Fermi systems, one faces the general problem that only a small fraction of atoms near the Fermi surface carries the temperature information, which reduces the detection sensitivity of imaging methods. For strongly interacting conditions, accurate thermometry is not straightforward and the interpretation of density profiles requires detailed knowledge of the equation of state \cite{Luo2007mot, Nascimbene2010ett, Ku2012rts} to extract temperature information from thermodynamic observables. For the specific case of a unitary Fermi gas the thermodynamics follows universal behavior \cite{Ho2004uto}, and thermometry is now well established, but this is not the case for the general situation of Fermi gases in strongly interacting regimes.

The working principle of thermometers in our daily life is to use a probe in thermal equilibrium with the object under investigation and to rely on a phenomenon with an easily detectable and well-understood temperature dependence. In our experiments, we adopted this principle by using a Bose gas of $^{41}$K to probe the Fermi sea of $^6$Li. This works either in the thermal impurity regime, with the temperature extracted from ballistic expansion, or in the condensed regime, where the BEC fraction provides us with a sensitive measurement of temperature, Here we summarize corresponding experiments carried out with our $^6$Li-$^{41}$K Fermi-Bose mixture and described in detail in Ref.~\cite{Lous2017toa}.

\begin{figure}
\centering
\includegraphics[width=0.85\linewidth]{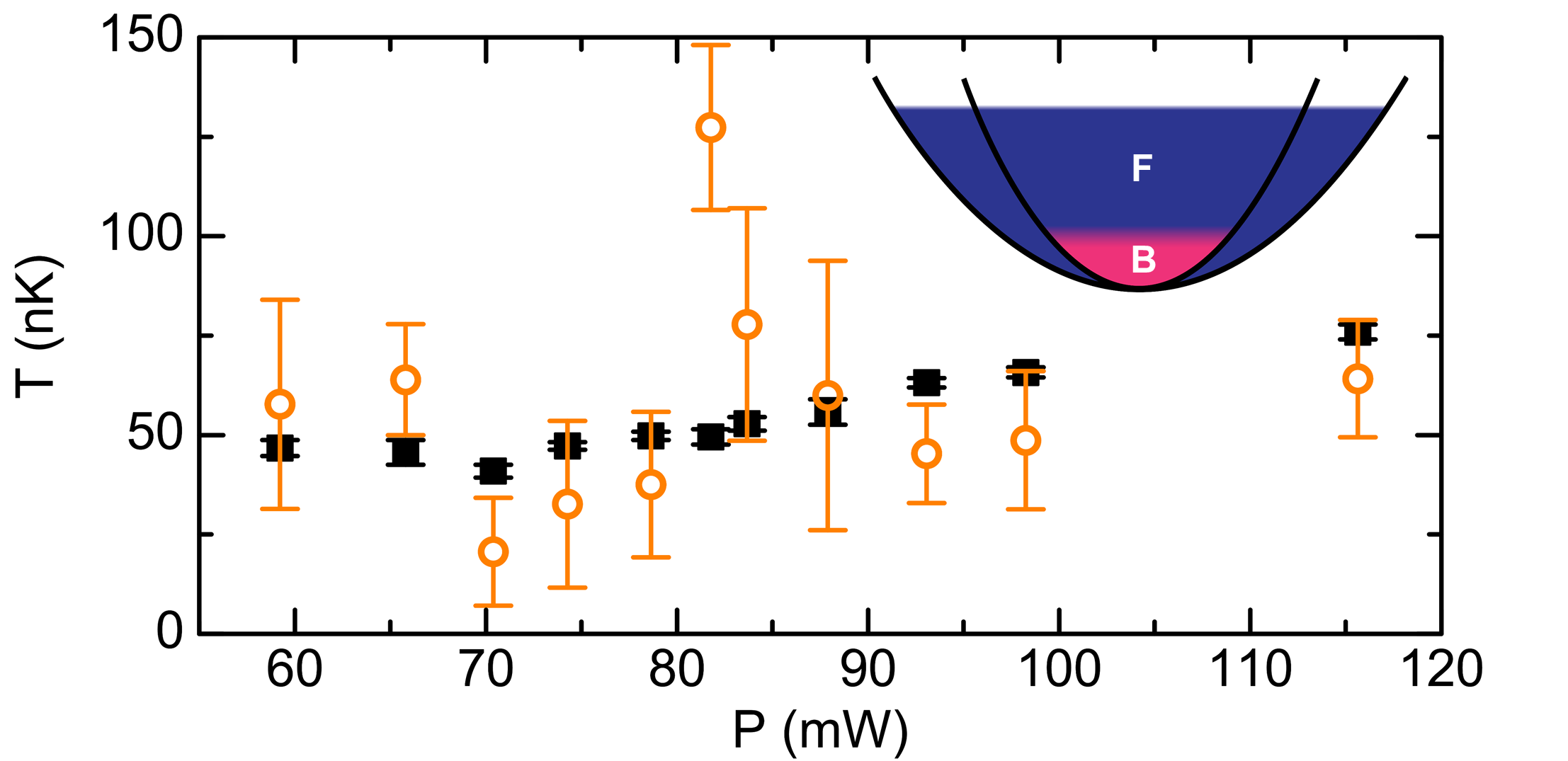}\\
\caption{Fermi gas thermometry based on a bosonic probe. The inset demonstrates the basic idea of a 
small sample of bosonic atoms (B) immersed in a large, deeply degenerate sea of fermions (F) under thermal equilibrium conditions. The harmonic trapping potentials (solid lines) are different for both species, depending on the particular trapping configurations used. 
The main panel shows a comparison of the method to obtain the temperature from the condensate fraction (filled symbols) 
with measurements based on the thermal component (open symbols), both recorded in time-of-flight images. The temperature is plotted versus the final power of the optical trap at the end of the evaporative cooling sequence.  Adapted from Ref.~\cite{Lous2017toa}.}
\label{fig:thermo}
\end{figure}

To explain the general principle of BEC thermometry, we consider a three-dimensional harmonic trap, which contains $N_b$ bosonic and $N_f$ fermionic atoms in thermal equilibrium, with both species having the same temperature $T$. The geometrically averaged trap frequencies are represented by $\bar{\omega}_b$ and $\bar{\omega}_f$. 
Combining textbook relations for the condensate fraction $\beta = 1-(T/T_c)^3$, the critical BEC temperature $k_B T_c = 0.940 \, \hbar \bar{\omega}_b \, N_b^{1/3}$, and the Fermi temperature $k_B T_F = 1.817 \, \hbar \bar{\omega}_f \, N_f^{1/3}$, one obtains the relation \cite{Lous2017toa}
\begin{equation}
    \frac{T}{T_F} = 0.518 \, (1-\beta)^{1/3} \, \frac{\bar{\omega}_b} {\bar{\omega}_f} \left( \frac{N_b}{N_f} \right)^{1/3} \, .
    \label{eq:BECthermo}
\end{equation}
This simple relation connects the reduced temperature $T/T_F$ with the condensate fraction $\beta$, which is a well-accessible observable in the experiments.
For typical trapping conditions,  
with $\bar{\omega}_b/\bar{\omega}_f \approx 0.5$ and $N_b/N_f \approx 1/30$, Eq.~(\ref{eq:BECthermo}) simplifies to $T/T_F \approx 0.083 \, (1-\beta)^{1/3}$. Assuming that we can measure the condensate fraction in the range
$0 \leq \beta \lesssim 0.95$, this translates into a temperature range of $0.03 \lesssim T/T_F \lesssim 0.08$, right in the interesting regime for experiments on deeply degenerate Fermi gases.

In our experiments on BEC thermometry \cite{Lous2017toa}, we used a single-beam optical dipole trap with axial confinement provided by the curvature of the magnetic field used for interaction tuning via Feshbach resonances ($\bar{\omega}_f/2\pi = 140\,$Hz after full evaporation). This scheme has proven a powerful tool for deep cooling of fermionic spin mixtures at high magnetic fields, see e.g.\ Ref.~\cite{Jochim2003bec, Bartenstein2004cfa}. The magnetic bias field was set to 1180\,G to achieve a large scattering cross section for elastic collisions between the different spin states of $^6$Li \cite{Varenna2006book, Bartenstein2005pdo, Zurn2013pco}. Note that, in contrast to all other Li-K experiments reported in these lecture notes, we do not exploit an interspecies Feshbach resonance, but rely on the background scattering length $a \approx 60\,a_0$ (Sec.~\ref{ssec:FR1}) for thermalization in the mixture.

In Fig.~\ref{fig:thermo}, we report a comparison between the temperature extracted from the BEC fraction (filled black symbols) and from the expansion of the thermal component of the K cloud (open orange symbols) after evaporative cooling. The temperatures were measured as a function of the final trap power $P$ at the end of the evaporative cooling process. We see that, while the extracted values are consistent with each other for the two methods, the temperatures derived from the BEC fraction show much smaller uncertainties. The comparison clearly demonstrates the superiority of the condensate fraction method in the temperature regime of interest.

The lowest temperatures that we observed in such measurements (at $P \approx 75\,$mW) correspond to $T/T_F \approx 0.059(5)$. This highlights the power of BEC thermometry for optimizing the cooling performance to achieve extremely cold Fermi gases in future experiments.

\subsubsection{Phase separation and Fermi-Bose interface}
\label{sssec:phasesep}

A macroscopic manifestation of strong repulsion between a Bose-Einstein condensate and a Fermi sea is the effect of phase separation.
The phenomenon has been studied extensively in degenerate Bose-Bose mixtures~\cite{Papp2008tmi, Tojo2010cps, Mccarron2011dsb, Stamperkurn2013sbg, Wacker2015tds, Wang2015ads, Lee2016psa}, where interactions are dominated by mean-field potential energies. The situation, however, becomes more complicated when fermionic constituents are involved, as strong repulsion on the scale of the Fermi energy is required to observe phase separation. 
Early experiments on Fermi-Bose mixtures of $^{40}$K and $^{87}$Rb~\cite{Zaccanti2006cot, Ospelkaus2006toh} have shown
qualitative signatures of phase separation. Our $^6$Li-$^{41}$K mixture has provided us with an interesting system for more in-depth studies related to this phenomenon.

In our experiments described in Ref.~\cite{Lous2018pti}, we probed the interface between the bosonic and the fermionic component in a phase-separated mixture, which consisted of a small BEC of $^{41}$K residing in a large Fermi sea of $^6$Li. We quantified the residual spatial overlap between the two components by measuring three-body recombination losses for variable strength of the interspecies repulsion. Our results demonstrate a corresponding smoothing of the transition between a separated and a non-separated mixture in a system of finite size.

\begin{figure}
\centering 
\includegraphics[width=0.9\textwidth]{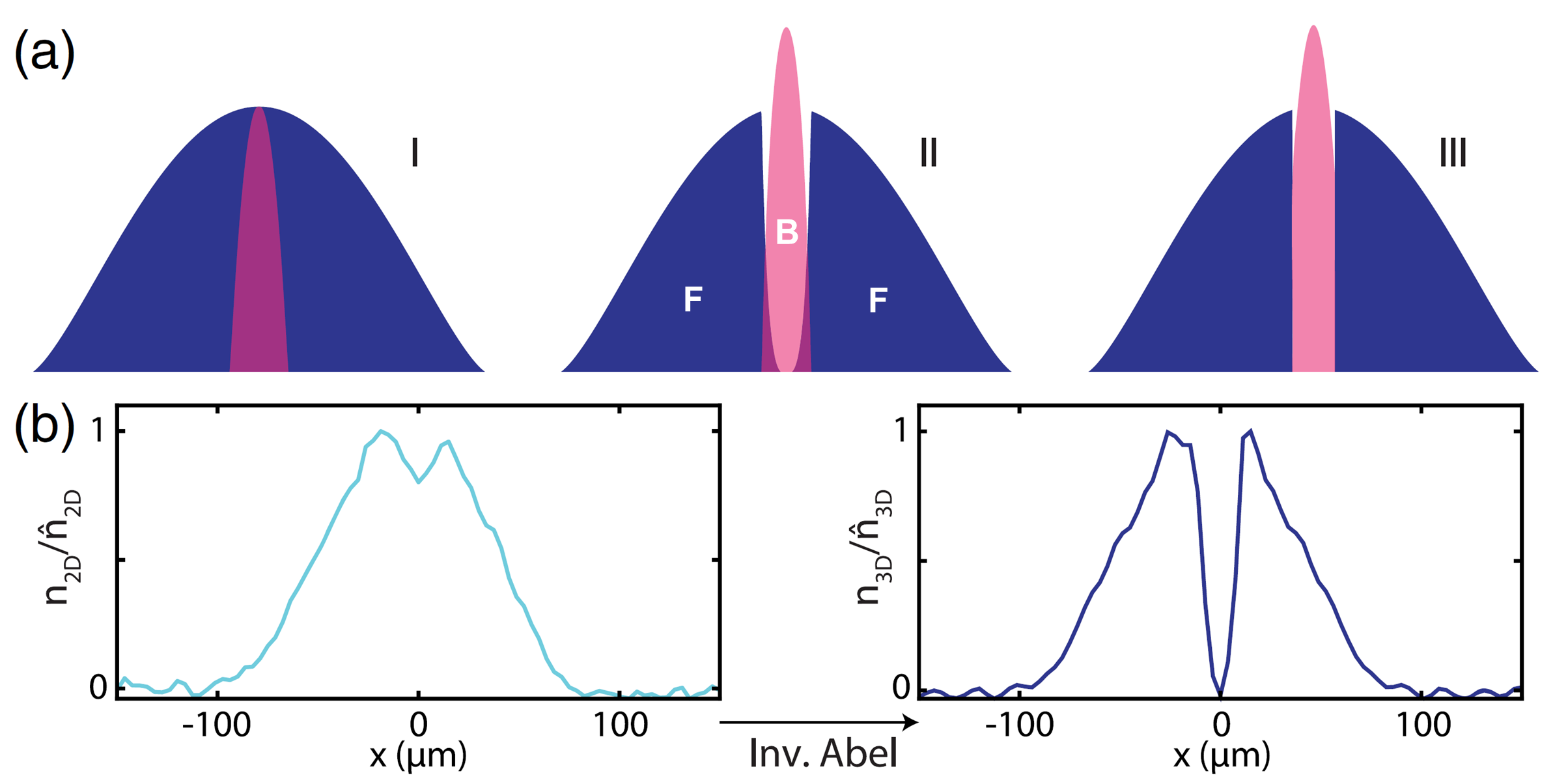}
\caption{Emergence of phase separation. (a) Schematic density profiles for fermions (F) and condensed bosons (B) for an increasing repulsive interaction. The densities are normalized to the corresponding peak value without an interaction. Note that in reality the boson peak density is a factor of 40 larger than the fermion peak density. (b) Experimentally observed normalized column density of a cut through the fermionic cloud and normalized reconstruction of the corresponding radial density profile using the inverse Abel transformation. Adapted from Ref.~\cite{Lous2018pti}.}
\label{fig:phasesep1}
\end{figure}

Figure~\ref{fig:phasesep1}(a) illustrates the onset of phase separation with increasing interspecies repulsion, showing the density profiles of a small-sized BEC coexisting with a large Fermi sea in a harmonic trap. For a vanishing interspecies interaction, the independent spatial profiles of the clouds show maximum overlap (I). With an increasing repulsion, the density of the lithium atoms in the center of the trap decreases, the BEC is compressed, and the spatial overlap between the clouds is reduced (II).  For strong repulsive interactions, the two clouds undergo phase separation (III), and the bosons reside at the center of the trap, forming a hole in the Fermi sea, as visible  in Fig.~\ref{fig:phasesep1}(b).

The main conditions and criteria for phase separation in such Bose-Fermi mixtures have been theoretically introduced in Refs.~\cite{Molmer1998bca, Viverit2000ztp, Roth2002sas}. Accordingly, we define the \textit{critical scattering length} 
\begin{equation}
    a_c = 1.15 \sqrt{a_{bb} / k_F} \, ,
    \label{eq:acrit}
\end{equation}
which marks the onset of phase separation in an infinitely large, homogeneous system. This quantity essentially results as the geometric average of two length scales~\footnote{This is similar to a two-component Bose-Bose mixture, where the relation of the geometric average of the two intracomponent scattering lengths to the   intercomponent scattering length determines whether the system is miscible or immiscible \cite{Lamporesi2022tcs}.}: the intraspecies scattering length $a_{bb}$ of the bosons and the inverse Fermi wavenumber $1/k_F = \hbar/\sqrt{2 m_f E_F}$, the latter characterizing the typical interparticle spacing in a degenerate Fermi gas. The prefactor ($1.15$ for our Li-K mixture) depends only weakly on the particular mass ratio of the mixture \cite{Huang2019bmo}.

The BEC is much smaller than the fermion cloud and occupies a very small volume within the Fermi sea. Thus, even strongest interaction with the BEC can cause only a local perturbation of the Fermi sea with negligible effect on the global scale. This scenario enables a description in terms of a \textit{fermionic reservoir approximation} (FRA), which assumes a homogeneous environement characterized by a constant Fermi temperature.

With increasing repulsion, phase separation emerges from a reduced overlap of both species, which finally results in the phase transition. In Ref.~\cite{Lous2018pti} we demonstrated a new method to characterize this overlap reduction. We used the effect of \textit{three-body recombination} as a probe, measuring corresponding losses from the mixture for a variable interspecies scattering length $a_{bf}$, which we tuned by means of our 335-G Feshbach resonance (Sec.~\ref{sssec:tune}).

The dominant loss process is a three-body collision involving two bosons and one fermion, and the local loss rate is thus proportional to $n_f n_b^2$, where $n_f$ and $n_b$ are the local number densities of the fermions and bosons, respectively. The total loss rate results proportional to the product of a loss rate coefficient $L_3$ and an overlap integral, both depending on the scattering length $a_{bf}$. To discuss the overlap effect separately, we define an \textit{overlap factor} 
\begin{equation}
    \Omega \equiv \frac{\int n_f n_b^2 dV}{\int \tilde{n}_f \tilde{n}_b^2 dV} \, ,
\end{equation}
with the $L_3$-dependence dropping out of the problem. Here $\tilde{n}_f$ and $\tilde{n}_b$ represent the number densities for vanishing interspecies interaction ($a_{bf}=0$), which can be easily calculated with textbook formulas for zero-temperature Fermi and Bose gases. The factor $\Omega$ provides a direct measure for the overlap reduction. In Ref.~\cite{Lous2018pti}, we also presented an extension of this approach to the finite-temperature case with a thermal bosonic component, defining an effective overlap factor $\Omega_{\rm eff}$.


Our experiments were performed in the crossed-beam optical dipole trap described in Sec.~\ref{sssec:ODT} with trap frequencies $\bar{\omega}_f/2\pi = 155\,$Hz and $\bar{\omega}_b/2\pi = 89\,$Hz. About $N_f = 1.3 \times 10^5$ and $N_b = 2.8 \times 10^4$ were prepared at temperature of $T = 87\,$nK. For the Fermi temperature of $T_F = 690\,$nK, this corresponds to $T/T_F = 0.13$. The condensate fraction of the K-BEC was around 0.5. With $1/k_F = 4570\,a_0$ and $a_{bb}=60.8\,a_0$, Eq.~(\ref{eq:acrit}) gives a moderate value of $a_c = 610\,a_0$ for the critical scattering length. 

\begin{figure}
    \centering
        \includegraphics[width=0.75\textwidth]{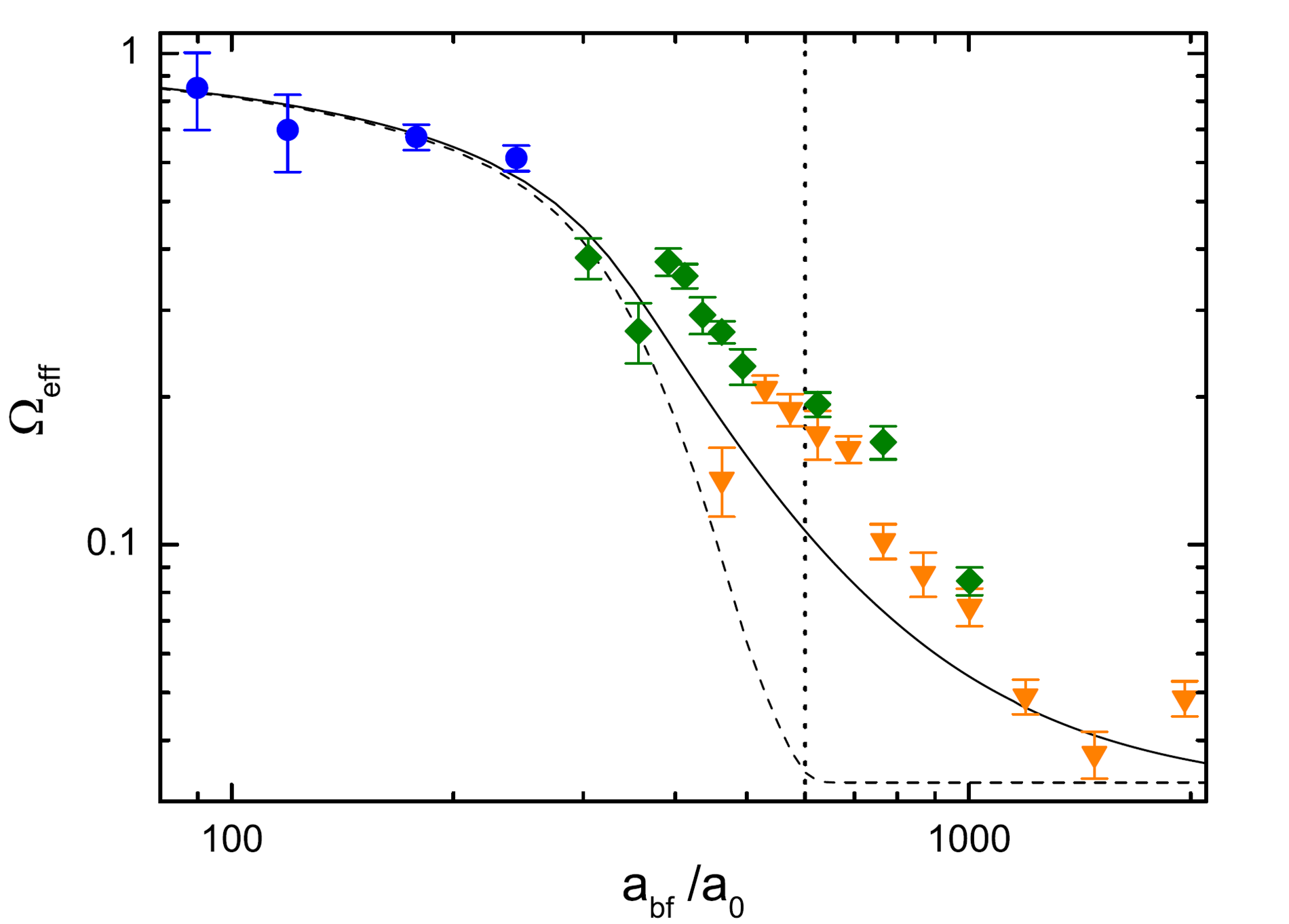}
\caption{Effective overlap factor versus interspecies scattering length. The measured values are shown by the symbols (inverse triangles, diamonds, and circles referring to three slightly different data sets \cite{Lous2018pti}). The error bars reflect the statistical uncertainties. The vertical dotted line marks the critical scattering length $a_c$, where phase separation is expected for a large homogeneous system. The solid line shows the results of a full numerical calculation (see text) and the dashed line our results obtained within the Thomas-Fermi approximation. From Ref.~\cite{Lous2018pti}.}
\label{fig:phasesep2} 
\end{figure}

Figure~\ref{fig:phasesep2} shows our measured values for $\Omega_\mathrm{eff}$ in comparison with theoretical models.  We qualitatively distinguish three regions. Below $a_{bf}\approx 250\,a_0$, the overlap reduction stays small. 
Then, as $a_{bf}$ further increases to about $1000\,a_0$, the spatial overlap drastically decreases to a small value of about $0.04$. 
For larger scattering lengths, $\Omega_\mathrm{eff}$ tends to remain at this small value. 
Contrary to the expectation of full phase separation at $a_{bf} \approx 600\,a_0$ (vertical dotted line in Fig.~\ref{fig:phasesep2}), we observe that beyond this point a considerable spatial overlap remains, which then decreases smoothly with further increasing scattering length. The observed behavior does not reveal any discontinuity related to a phase transition. 

To interpret the observed behavior of $\Omega_\mathrm{eff}$, we constructed a numerical mean-field model~\footnote{See Supplementary Material of Ref.~\cite{Lous2018pti} and Ref.~\cite{Huang2020bec}.}, which allows us to calculate the density distributions for an interacting Bose-Fermi mixture at a zero temperature for our typical experimental parameters. Our model starts from the energy functional of the mixture as given by Refs.~\cite{Imambekov2006bot,Trappe2016gsd}, and we use imaginary time evolution to vary the BEC and the fermionic densities and to minimize the energy functional. At the end, the evolution gives the static solution of $n_{f}$ and $n_{b}$ at zero temperature. Since we have a partial BEC, we additionally take into account the thermal bosonic density $n_{t}$. 
With these density distributions, we numerically calculate the overlap integrals and the effective overlap factor $\Omega_\mathrm{eff}$.

The results of our numerical model are represented by the dashed and solid curves in Fig.~\ref{fig:phasesep2}. For the dashed curve, the densities are obtained within the Thomas-Fermi approximation, which neglects the kinetic energy of the spatially varying BEC. The results indeed show a rapid decrease of $\Omega_\mathrm{eff}$ until full phase separation sets in at about $600\,a_0$. Then, in a fully phase-separated regime, a plateau is reached where only the thermal bosonic component can lead to losses. Evidently, this theoretical behavior is not consistent with the experimental data points.  A notably smoother decrease of $\Omega_\mathrm{eff}$ results from our numerical model (solid line in Fig.~\ref{fig:phasesep2}), when we consider the full energy functional which includes the kinetic energy of the BEC as well as the much weaker density gradient correction from the Fermi gas~\cite{Imambekov2006bot}. Within the residual uncertainties of our method, this model reproduces the observed behavior very well.

Our results show that the kinetic energy term prevents the BEC density from changing abruptly. This plays an essential role in smoothing the density profiles of the separated components near the interface and, thus, in maintaining the residual spatial overlap. Accordingly, the relevant length scale that determines the thickness of the interface layer corresponds to the BEC healing length~\cite{Dalfovo1999tob}, which for our present conditions can be estimated to $\sim$$0.50\,\mu$m. This length scale can be compared with the shortest macroscopic length scale of the system, which in our case is the radial size of the BEC of a few micrometers. 
The measured overlap factor can be understood as the volume ratio of the interface layer and the whole BEC, and the smoothing of the phase transition can thus be interpreted as a consequence of the \textit{finite size} of the system~\cite{Binder1984fss, Brezin1985fse}.

\subsubsection{Phase separation and BEC breathing mode}
\label{sssec:breathing}

Further insight into the phase-separated state can be obtained by investigating the \textit{dynamic behavior} of the BEC, which is compressed by the Fermi sea. In Ref.~\cite{Huang2019bmo} we studied the radial breathing mode and found drastic upshifts of its frequency, depending on the interaction strength and on the atom number ratio of bosons and fermions.

The frequency shift can be understood qualitatively by
considering the interface that emerges from phase separation
of the Bose-Fermi mixture. In the presence of the interface,
the BEC becomes hydrostatically compressed by the Fermi
pressure. Exciting a collective mode of the BEC leads to a
motion of this interface. If the mode is a breathing mode, the
oscillation inflates and deflates the interface, like modulating a bubble in the Fermi sea. Intuitively, the volume change of the
BEC leads to a significant reversible work against the Fermi
pressure. Because of the existence of this strong restoring
mechanism the oscillation frequency substantially increases.
This stands in contrast to surface modes of BECs immersed
in Fermi gases, which have been observed in experiments
\cite{DeSalvo2019oof, Ferrierbarbut2014amo, Delehaye2015cva, Roy2017tem, Wu2018cdo}. There, the frequency shifts are rather small, since surface modes do not change the volume and thus do no work
against the Fermi pressure.

\begin{figure}
\centering
\includegraphics[width=0.75\linewidth]{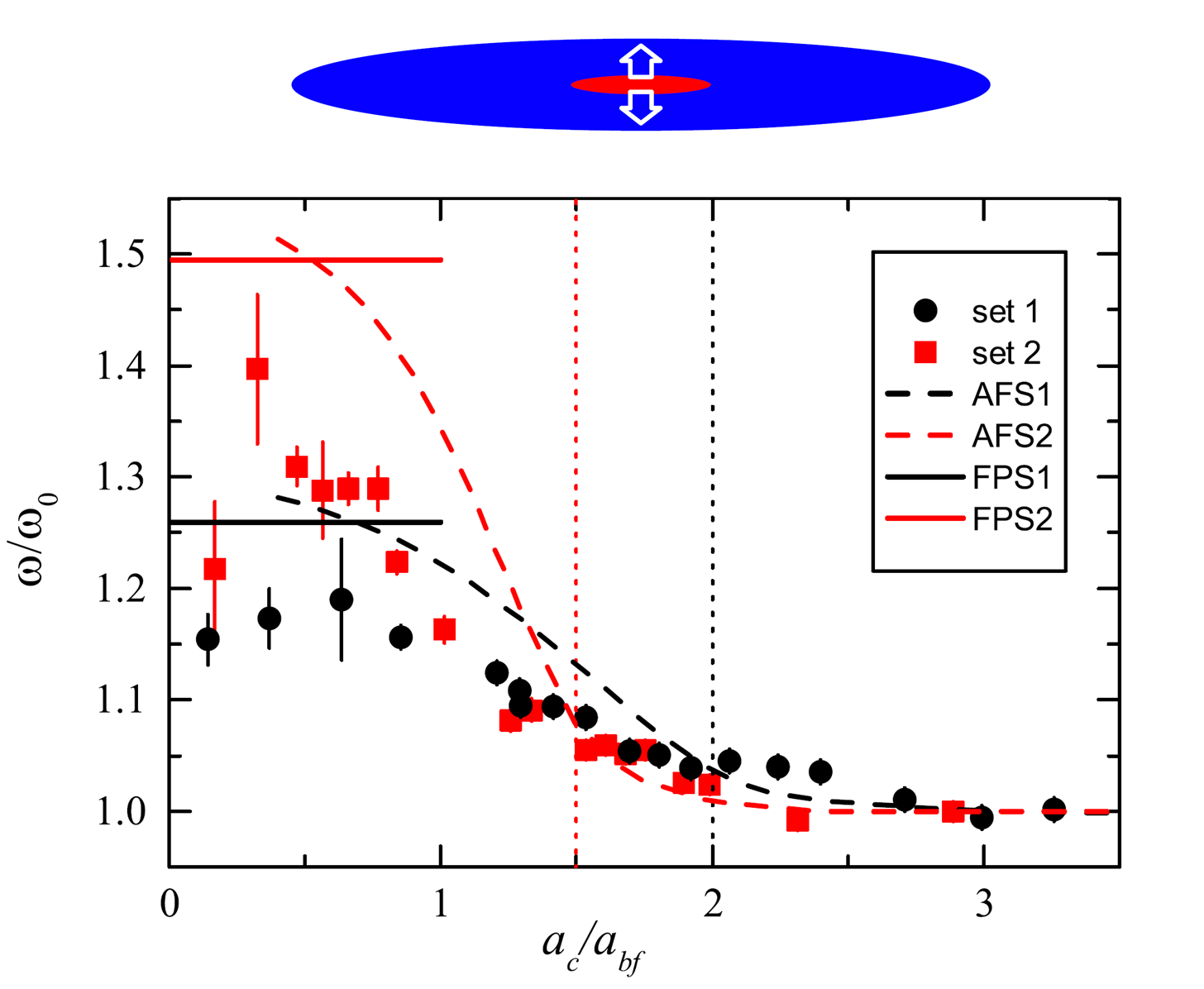}
\caption{Breathing mode frequency in the strongly repulsive Fermi-Bose mixture of $^6$Li and $^{41}$K. The radial breathing mode of the cigar-shaped cloud is illustrated on top of the figure. The data points, plotted as a function of the dimensionless parameter $a_c/a_{bf}$ refer to two different sets of measurements, taken at  number ratios $N_b/N_f = 0.16$ (set~1) and $N_b/N_f = 0.05$ (set~2). The vertical dashed lines mark the corresponding values of $a_c/a_d$, where the depletion of the central fermion density sets in. The experimental results are compared with two different models (see text).
Adapted from Ref.~\cite{Huang2019bmo}. }
\label{fig:phasesep3}
\end{figure}

To excite the breathing mode of the K condensate we modulated the interspecies interaction by periodically changing the scattering length~\cite{Matthews1998dro, Pollack2010ceo}. After a variable hold time in the trap, we released the cloud from the trap and took time-of-flight images of the expanding cloud. To obtain the frequency $\omega$ of the breathing mode, we fitted the recorded time evolution of the width of the BEC with a damped harmonic oscillation. For reference, we also measured the corresponding value $\omega_0$ of the bare oscillation for vanishing interspecies interaction ($a_{bf}=0$).

In Fig.~\ref{fig:phasesep3} we present our measurements of the normalized breathing mode frequency $\omega/\omega_0$ as a
function of the dimensionless interaction parameter $a_c/a_{bf}$
with the critical scattering length $a_c$ given by Eq.~(\ref{eq:acrit}).  We show two sets of measurements, taken with different atom number ratios $N_b/N_f$. Both sets show essentially the same behavior. With increasing $a_{bf}$ the frequency shift begins to appear close to a \textit{depletion scattering length} $a_d$, which is defined as the value where the central number density of the fermions is reduced to zero~\cite{Huang2019bmo}.
Then, in the intermediate range of
$a_c/a_{bf}$ between 2 and 1,  $\omega/\omega_0$ rapidly rises until a plateau value of about 1.2 is reached for $N_b/N_f = 0.16$ (1.3 for $N_b/N_f = 0.05$). For even stronger repulsion
in the phase-separated regime, no further frequency change
is observed. These results show that the frequency upshift
emerges exactly where the transition to the phase-separated
regime occurs and finally levels off at the plateau value when
full phase separation is reached. Quantitatively, this behavior depends slightly on the boson-fermion number ratio. For $N_b/N_f = 0.05$ the observed frequency changed was as large as 30\%.

The theoretical description of the observed dynamics is challenging because of the intricate collective dynamics in the excited Fermi sea and because of beyond-mean-field effects at strong repulsive coupling. Different models have been introduced in Refs.~\cite{Huang2019bmo, Huang2020bec, Grochowski2020bmo}, which capture the observed behavior. The most intuitive description is the \textit{adiabatic Fermi sea} (AFS) model, which assumes that the Fermi sea follows the BEC oscillation in a quasi-static equilibrium. As shown by the dashed lines in Fig.~\ref{fig:phasesep3}, this model captures the main behavior qualitatively in the regime where the Fermi sea is not completely depleted in the trap center, but it overestimates the maximum frequency shift near resonance. An alternative model, which considers the regime of \textit{full phase separation} (FPS) and the full dynamic response of the Fermi sea, gives better agreement near resonance. Further insight is provided by Ref.~\cite{Grochowski2020bmo} based on a hydrodynamic description of the interface (domain wall) between bosons and fermions, and the inclusion of finite-temperature effects.

\section{Mixtures of Dy and K: Towards novel superfluids}
\label{sec:DyK}

With strongly magnetic lanthanide atoms of dysprosium and erbium entering the stage of quantum-degenerate gases \cite{Lu2011sdb, Aikawa2012bec}, new opportunities also emerged for experiments with fermions. It turned out that evaporative cooling of the fermionic isotopes $^{167}$Er and $^{161}$Dy can be very efficient, even in fully spin-polarized samples \cite{Ravensbergen2018poa, Lu2012qdd, Aikawa2014rfd}. This also boosted the interest in exotic mixtures involving these fermionic species. In the mid 2010's, driven by the prospects of creating novel strongly interacting fermion mixtures and superfluid states, we started a new line of experimental research on mixtures of $^{161}$Dy and $^{40}$K.

The reason for our choice of this Fermi-Fermi mixture can be explained based on three criteria: 
First, the mass ratio should be in an intermediate range, much larger than $1$, but staying well below a critical value of $13.6$ to suppress Efimov-related losses~\cite{Marcelis2008cpo}. 
Second, there should be the prospect of interaction tuning by sufficiently broad Feshbach resonances, which excludes atoms with a closed-shell structure from our menu~\footnote{Combinations of a closed-shell fermion ($^{171}$Yb, $^{173}$Yb, or $^{87}$Sr) in its electronic ground state with an alkali-metal atom offer only extremely narrow resonances \cite{Zuchowski2010urm, Barbe2018oof}.}. 
Third, on a practical side, the atoms should be easy to handle in experiments~\footnote{This also excludes the $^{163}$Dy isotope because of complications in the laser cooling and preparation processs owing to the normal (not inverted) ground-state hyperfine structure~\cite{Lu2012qdd}.}. Complying to these criteria narrows down the possible combinations to $^{161}$Dy-$^{40}$K,  $^{167}$Er-$^{40}$K and $^{53}$Cr-$^{6}$Li, from which we selected the first combinations. While the second one (Er-K) has, to our best knowledge, not been realized so far, the Cr-Li system is under investigation in Florence (see M.~Zaccanti's contribution to these proceedings \cite{Zaccanti2022mif}).


In the following, we describe the main advances in our Dy-K experiments.
We start with a summary of the
main procedures to prepare the double-degenerate mixture (Sec.~\ref{ssec:DyKprepare}). The tunability of the mixture via Feshbach resonances is then discussed (Sec.~\ref{ssec:DyKFR}), before we present an experimental breakthrough on the creation of a resonant mixture and its hydrodynamic behavior (Sec.~\ref{ssec:DyKhydro}). Finally, we demonstrate the creation of bosonic $^{161}$Dy$^{40}$K Feshbach molecules and the production of a trapped, near-degenerate molecular sample (Sec.~\ref{ssec:DyKmols}).

\subsection{Preparation of a double-degenerate mixture}
\label{ssec:DyKprepare}

The preparation of the mixture is described in detail in Ref.~\cite{Ravensbergen2018poa}. Here we provide a brief summary of the main points.

Using standard laser cooling techniques, the two species are prepared in magneto-optical traps and loaded sequentially into a large-volume optical dipole trap. Doppler cooling of \Dy on the narrow intercombination line at a wavelength of 626\,nm \cite{Maier2014nlm, Dreon2017oca}, and gray-molasses cooling of \K \cite{Fernandes2012sdl} provide similar temperatures around 10\,$\mu$K. Both species are pumped into their lowest internal states ($F=21/2$, $m_F=-21/2$ for Dy, and $F=9/2$, $m_F=-9/2$ for K), which happens naturally in the Dy MOT \cite{Dreon2017oca} and needs additional Zeeman pumping for the K sample. The polarized mixture in the large-volume dipole trap then serves as a reservoir for further evaporative cooling.

\begin{figure}
\centering 
\includegraphics[width=0.9\textwidth]{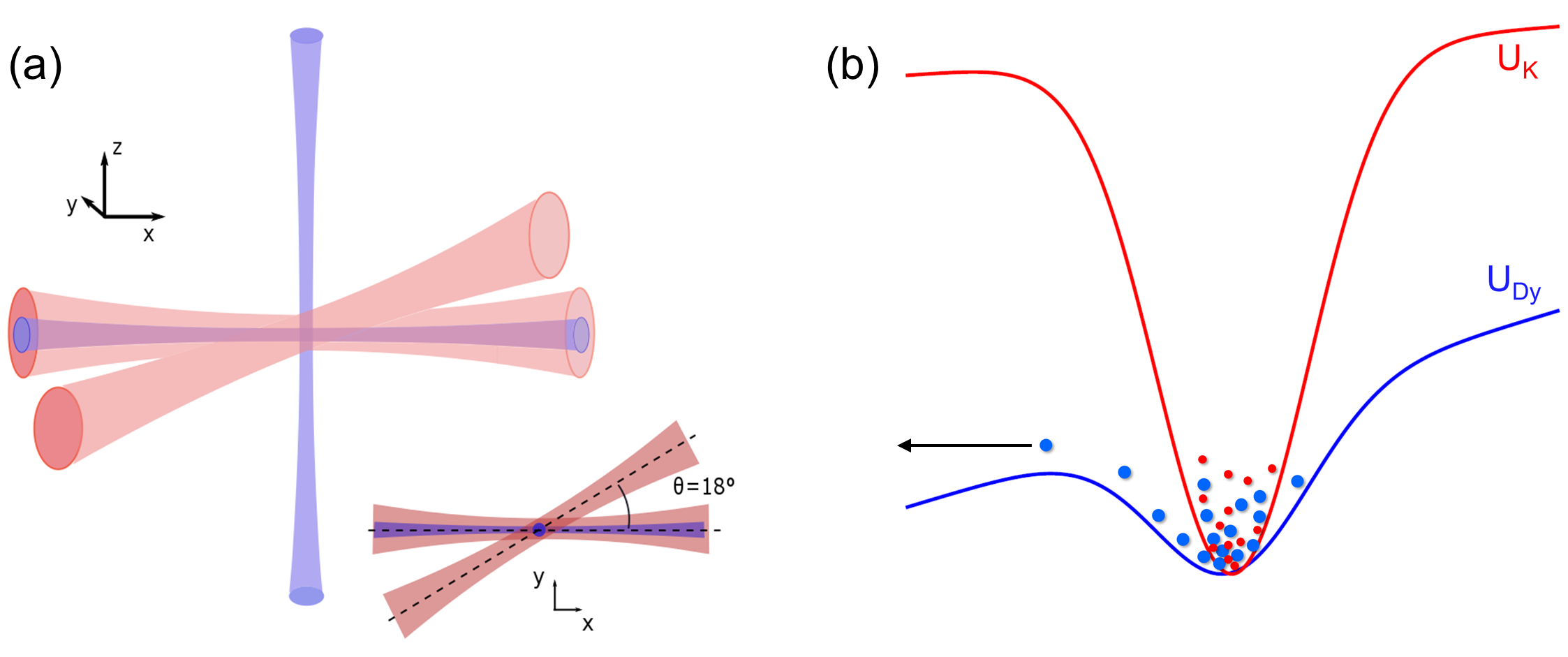}
\caption{Optical dipole trapping and evaporative cooling of the Dy-K Fermi-Fermi mixture. (a) The two wider beams crossing in the  horizontal $x$-$y$ plane under an angle of $18^{\circ}$ constitute the large-volume reservoir trap, which is loaded with laser-cooled atoms from magneto-optical traps. The two tighter beams in the $x$- and $z$-directions form the compressed trap used for evaporative cooling. 
(b) Illustration of evaporative cooling. The trapping potential of Dy is shallower because of the 3.2 times lower dynamic polarizability at the trap wavelength of 1064\,nm, and because of the effect of gravity. The K atoms thus do not evaporate and they are cooled sympathetically.
Adapted from Ref.~\cite{Ravensbergen2018poa}.}
\label{fig:DyKtrap}
\end{figure}

For evaporative cooling, we transfer the mixture into a much tighter crossed-beam optical dipole trap, which overlaps with the reservoir trap, see illustration in Fig.~\ref{fig:DyKtrap}(a). The evaporation sequence then needs about 15\,s, in which the trap power is slowly ramped down. It is important to note that the optical potential is by a factor of about $3.2$ deeper for K than for Dy, owing to the different dynamic polarizabilities at the dipole trap wavelength \cite{Ravensbergen2018ado}. As shown in Fig.~\ref{fig:DyKtrap}(b), Dy is evaporated out of the trap, while K stays in the trap and is cooled sympathetically. 
This cooling scheme is closely related to earlier experiments on the attainment of quantum degeneracy in Yb-Li mixtures \cite{Hara2011qdm} and on Yb-Rb mixtures \cite{Hansen2011qdm, Hansen2013poq}, where the heavy lanthanide species experiences a shallower trapping potential than the light alkali species.
In contrast to these experiments, however,
the evaporation process of Dy relies solely on universal dipolar scattering \cite{Bohn2009qud}, which allows to cool fully spin-polarized Fermi gases to quantum degeneracy \cite{Lu2012qdd, Aikawa2014rfd, Burdick2016lls}. For evaporative cooling we choose a magnetic bias field of 230\,mG, which minimizes three-body losses in Dy in the presence of a very dense spectrum of intraspecies Feshbach resonances \cite{Burdick2016lls, Soave2022lff}. At this magnetic bias field, the interspecies scattering length, estimated to about $40\,a_0$ \cite{Ravensbergen2018poa}, is sufficient for cross-species thermalization and thus sympathetic cooling of K by Dy.

The effect of gravity, different for both species, also plays an important role in the evaporative cooling of the mixture. The corresponding reduction of the trapping potential for the heavier species, see Fig.~\ref{fig:DyKtrap}(b), can further enhance the efficiency of the cooling process \cite{Hung2008aec}. The effect of gravity can be compensated by a magnetic `levitation' gradient \cite{Weber2003bec}. Because of the different atom masses and magnetic moments, levitation acts in a highly species-selective way \cite{Lous2017toa} and can serve as an important tool in the experiments, e.g.\ for a Stern-Gerlach separation of atoms and molecules \cite{Herbig2003poa}, see also Sec.~\ref{ssec:DyKmols}.

\begin{figure}
\centering 
\includegraphics[width=1.0\textwidth]{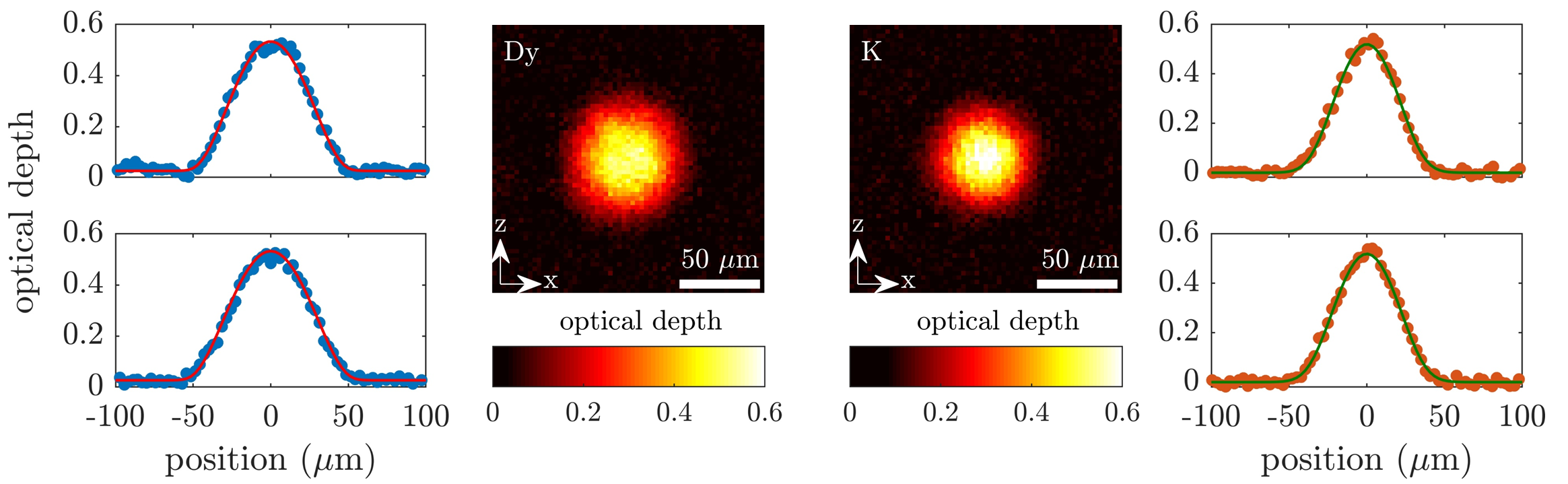}
\caption{Deep cooling of the Dy-K Fermi-Fermi mixture. 
Density profiles of $^{161}$Dy atoms after time-of-flight expansion of $7.5\,$ms and $^{40}$K atoms after $2.5\,$ms demonstrate reduced temperatures of $T_{\mathrm{Dy}}/T^{\mathrm{Dy}}_F \approx 0.1$ and 
$T_{\mathrm{K}}/T^{\mathrm{K}}_{F} \approx 0.2$, respectively. 
Here, the atom numbers are $N_{\mathrm{Dy}} \approx 2 \times10^{4}$ and  $N_{\mathrm{K}}\approx 4\times10^{3}$.
The absorption images (center) are shown together with cuts of the column density along the $x$ (top) and $z$ (bottom) axes  (left panel for Dy and right panel for K). The solid lines represent fits with a polylogarithmic function~\cite{DeMarco2001PhD}.
Adapted from Ref.~\cite{Ravensbergen2018poa}.}
\label{fig:DyKprepare}
\end{figure}

In Fig.~\ref{fig:DyKprepare}, we show a result of deep evaporative cooling from Ref.~\cite{Ravensbergen2018poa}. With some further improvements of the set-up, in particular the implementation of a in-trap cooling stage based on the very narrow Dy line at 741\,nm \cite{Ye2022ool}, we now reach typical working conditions
with $N_{\rm Dy} \approx 10^5$ at  $T_{\mathrm{Dy}}/T^{\mathrm{Dy}}_F \approx 0.13$ in a trap with $\bar{\omega}_{\rm Dy} = 2\pi \times 150\,$Hz. The number of K atoms in the tighter trap ($\bar{\omega}_{\rm K}/ \bar{\omega}_{\rm Dy} =3.60$ \cite{Ravensbergen2018ado}) is about a factor of 5 less than the number of Dy atoms, depending on the particular loading conditions. This also constitutes a significant improvement provided by the 741-nm narrow-line cooling.

While we perform evaporative cooling in the tight trap at a magnetic field of 230\,mG, subsequent experiments near interspecies Feshbach
resonances require a transfer of the mixture to higher magnetic fields. 
The system then inevitably has to cross many interspecies Dy-K resonances and intraspecies Dy resonances, the latter exhibiting an extremely large density of narrow Feshbach resonances (typically 50 resonances per gauss \cite{Soave2022lff}). To minimize losses and heating we decompress the sample by loading it adiabatically into a very shallow optical dipole trap with $\bar{\omega}_{\rm Dy}/2\pi \approx 40\,$Hz before we quickly (within a few ms) ramp up the magnetic field to its target value, where further experiments are carried out.

\subsection{Feshbach resonances and interaction tuning}
\label{ssec:DyKFR}

When starting to work with a yet unexplored atomic mixture, one cannot build on any a priori knowledge of the scattering properties. Thorough experimental investigations are needed to characterize the system and to identify Feshbach resonances for controlled interaction tuning. For mixtures involving strongly magnetic lanthanide atoms, scattering models that describe the complete two-body interaction physics are not available because of the enormous complexity of the system. Here the characterization of resonance features has to rely solely on experimental methods.

The arduous task of identifying and characterizing Feshbach resonances in the $^{161}$Dy-$^{40}$K mixture took a few years. We scanned a magnetic field range up to about 300\,G and observed many narrow interspecies resonances and even more Dy intraspecies resonances. To make this very long story short, we now discuss our main findings on two Feshbach resonances that we identified as particularly useful for interaction tuning.

In Ref.~\cite{Ravensbergen2018poa}, we characterized a rather broad resonance located near 217\,G, which is part of a scenario of three overlapping resonances. Identifying the locations of the poles and zero crossings of these resonances by measuring the rate of interspecies thermalization and applying the model of Ref.~\cite{Mosk2001mou} to derive the scattering cross section from thermalization rates, we obtained the resonance parameters (see definitions in Sec.~\ref{sec:FR}): $B_0=217.27(15)\,$G for the resonance  center, $A=1450(230)\,$G for the strength parameter, and $a_{\rm bg} = +59(9)\,a_0$ for the background scattering length. In later experiments  \cite{Soave2022PhD}, we also carried out measurements of the binding energy of the underlying molecular state. For the range parameter of the broad resonance we obtained the upper limit $R^*<150\,a_0$, which indicates universal behavior over a considerable part of the resonance width. Indeed, we have observed long lifetimes of the resonant mixture, which we attribute to the famous Pauli suppression effect of few-body losses in the two-component fermion mixture~\cite{Petrov2004wbd, Petrov2005spo, Petrov2005dmi}.
This makes the 217-G resonance a very interesting tool for experiments. We note, however, that there is also the disadvantage of the broad resonance to be contaminated by many narrow inter- and intraspecies resonances, which can cause losses and heating, in  particular when the magnetic field is ramped across.

\begin{figure}
\centering 
\includegraphics[width=0.95\textwidth]{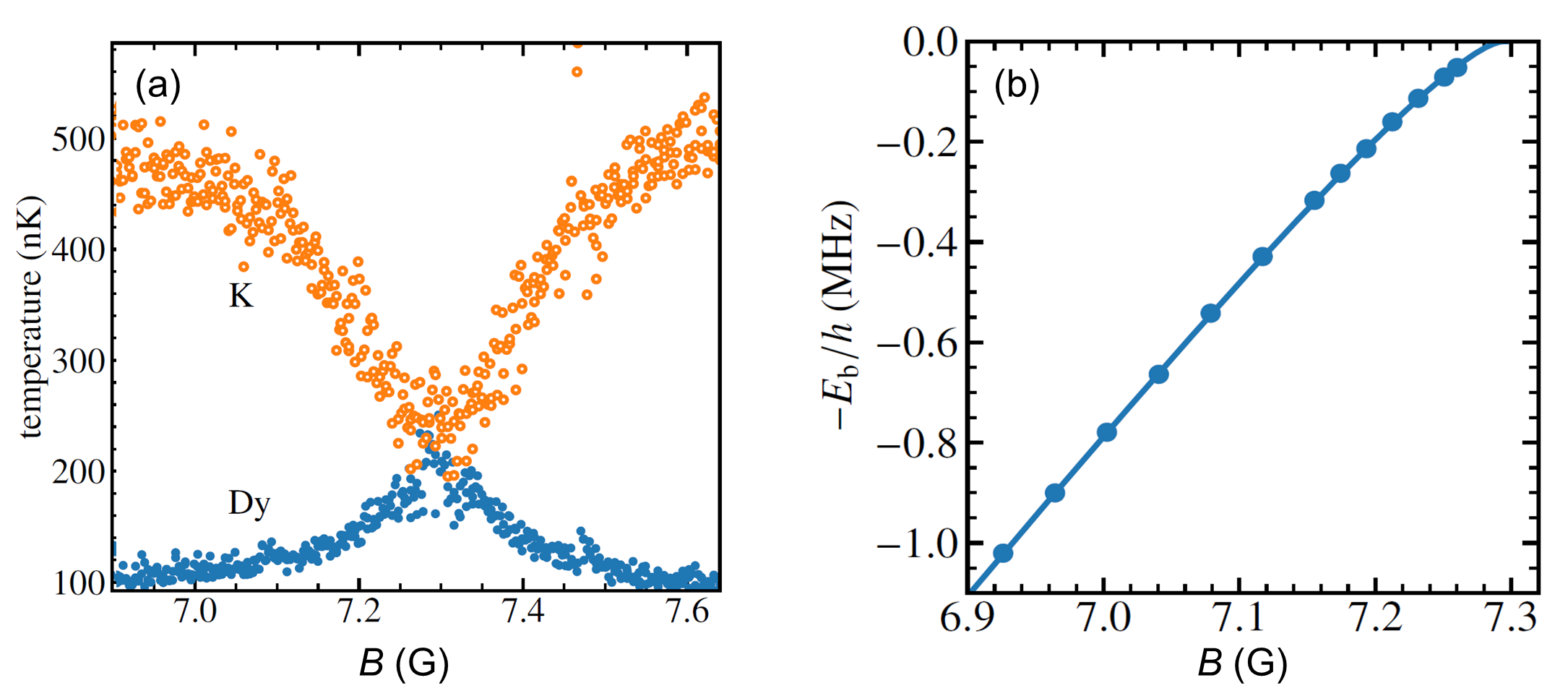}
\caption{Characterization of the 7.3-G Feshbach resonance by (a)  interspecies thermalization and (b) molecular binding energy measurements.
Adapted from Ref.~\cite{Ye2022ool}.}
\label{fig:DyK7p3}
\end{figure}

In Ref.~\cite{Ye2022ool}, we focused on the low-field region and identified a couple of further resonances of interest. The most promising one is located at 7.3\,G, at a low field very convenient for the experiments. In Fig.~\ref{fig:DyK7p3}, we illustrate two ways to characterize the resonance. In Fig.~\ref{fig:DyK7p3}(a), we show the results of an interspecies thermalization measurement. After some evaporative cooling, which brought the sample in the near-degenerate regime, we applied species-selective parametric heating to the K component, increasing its temperature to about 500\,nK while leaving the Dy component at 100\,nK. After a fixed hold time of 100\,ms at a variable magnetic field, we measured the temperatures of the two components. One clearly sees that the resonance leads to a fast thermalization. From the measured thermalization rates, values for the elastic scattering cross can be derived~\cite{Mosk2001mou} and thus for the modulus of the scattering length.            
In Fig.~\ref{fig:DyK7p3}(b), we show measurements of the binding energies of the underlying molecular state by magnetic-field modulation spectroscopy \cite{Chin2010fri, Claussen2003vhp}. A fit of Eq.~(\ref{eq:Eb}) to the experimental data yields the resonance parameters. The most accurate determination of these parameters to date can be found in Ref.~\cite{Soave2023otf}. Here binding energy measurements in combination with magnetic-moment spectroscopy \cite{Mark2007sou} yielded: $B_0 = 7.276(2)\,$G, $A = 24.0(6)\,$G, and $R^* = 604(20)\,a_0$.

Although being about $50$ times narrower than the broad 217-G resonance, the 7.3-G resonance offers important advantages. It is well-isolated from other interspecies resonances, and the region of interest close to the center is not contaminated by intraspecies Dy resonances. On the practical side, it is much easier to control the magnetic field in a fast and accurate way. Even with the larger value of $R^*$ and the smaller universal range, one can reach regimes of $R^*/a < 1$, where inelastic scattering processes of weakly bound dimers are suppressed \cite{Jag2016lof, Levinsen2011ada}. For controlled interaction tuning and, in particular, for experiments involving weakly bound Feshbach dimers (Sec.~\ref{ssec:DyKmols}), the 7.3-G resonance serves  us as a powerful tool.

\subsection{Hydrodynamic expansion of a resonant atomic mixture}
\label{ssec:DyKhydro}

Hydrodynamic properties are at the heart of the macroscopic behavior of strongly interacting Fermi gases \cite{Varenna2006book, Giorgini2008tou} and their understanding has challenged us from the very beginning of experiments in this field \cite{Ohara2002ooa}. A key question is to distinguish whether hydrodynamic behavior originates from collisional dynamics or superfluidity. 

In the Fermi-Fermi mixture of $^6$Li and $^{40}$K (Sec.~\ref{sec:LiK}), hydrodynamic behavior was observed in the expansion of the cloud \cite{Trenkwalder2011heo}, but the limited lifetime of this mixture has not allowed further studies. Our mixture of $^{161}$Dy and $^{40}$K offers new opportunities to study hydrodynamic behavior in a variety of different situations. Here, as a striking example, we summarize an experiment \cite{Ravensbergen2020rif} that demonstrates the locked expansion of the two fermionic components.

\begin{figure}
\centering 
\includegraphics[width=0.60\textwidth]{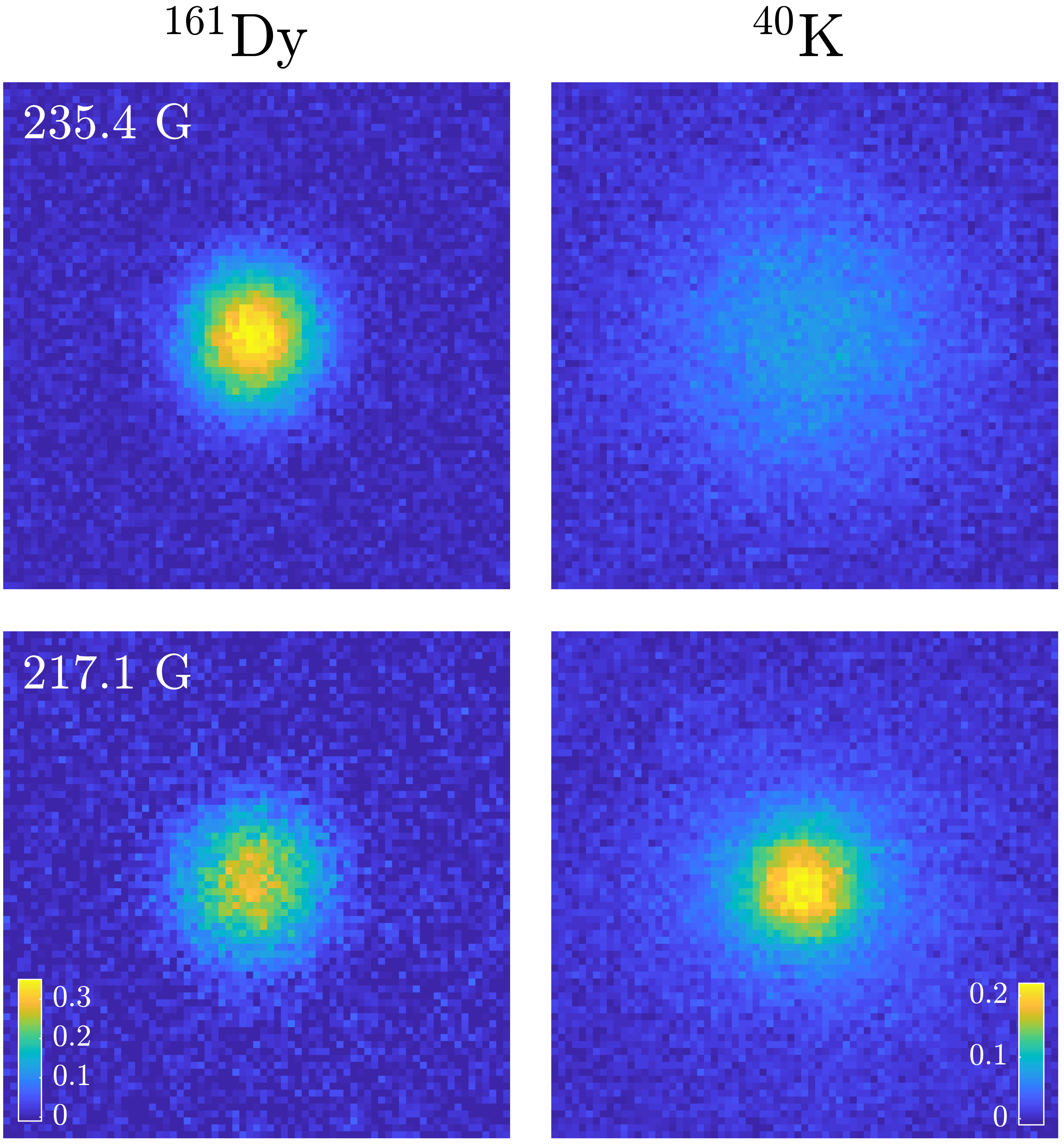}
\vspace{2mm}
\caption{Comparison of the expansion of the mixture for weak (upper row) and resonant (lower row) interspecies interaction. The absorption images show the optical depth for both species (Dy left, K right) after a time of flight of 4.5\,ms. The field of view of all images is $240\,\mu{\rm m} \times 240\,\mu$m.
From Ref.~\cite{Ravensbergen2020rif}.}
\label{fig:DyKhydro}
\end{figure}

The mixture of $N_{\rm Dy} = 2 \times 10^4$ and  $N_{\rm K} = 8 \times 10^3$  atoms was prepared in a trap with frequencies $\bar{\omega}_{\rm Dy}/2\pi = 120\,$Hz and  $\bar{\omega}_{\rm K}/2\pi = 430\,$Hz. The temperature corresponded to thermal conditions close to degeneracy with $T/T_F^{\rm Dy} = 1.7$ and $T/T_F^{\rm K} = 0.65$.  
The sample was released from the trap right after switching to the target magnetic field strength, and finally absorption images were taken. 

The absorption images in the upper row of Fig.~\ref{fig:DyKhydro} illustrate the case of weak interactions ($a \approx -40\,a_0$), realized at $B=235.4$\,G. Here the expansion takes place in a ballistic way and, as expected from the mass ratio, the K component expands much faster than the Dy component. In contrast, in the resonant case (images in the lower row of Fig.~\ref{fig:DyKhydro}) both components expand with similar sizes. Evidently, the interaction between the two species slows down the expansion of the lighter species and accelerates the expansion of the heavier species. Such a behavior requires many elastic collisions (for our experimental conditions we estimate a rate of $\sim$$10^4\,$s$^{-1}$) on the timescale of the expansion and can thus be interpreted as a hallmark of hydrodynamic behavior.  

A closer inspection of the spatial profiles of the hydrodynamically
expanding mixture revealed an interesting difference between the heavy and the light species. While the Dy cloud essentially kept its
near-Gaussian shape, the K cloud (initially about twice
smaller than the Dy cloud) developed pronounced side
wings \cite{Ravensbergen2020rif}. Apparently, the mixture formed a hydrodynamic core surrounded by a larger cloud of ballistically expanding lighter atoms.
The physical mechanism for the formation of the
latter (as confirmed by Monte Carlo simulations \cite{Ravensbergen2020rif, Kreyer2023PhD}) is the faster diffusion of lighter atoms, which can leak
out of the core and, in the absence of the other species,
begin to move ballistically. We point out that this bimodality
effect is not an experimental imperfection, but a
generic feature in the hydrodynamic expansion of a mass-imbalanced
mixture.

Similar bimodal expansion behavior was observed before in fermionic spin mixtures with a superfluid core~\cite{Zwierlein2006fsw}. Here the bimodality of the minority component constitutes a clear signature of superfluidity. However, such a conclusions is not possible for mass-imbalanced mixtures, where collisional hydrodynamics can produce such an effect.

More insight on the hydrodynamic dynamics of the mixture can be gained by studying collective oscillation modes, such as the ``species-dipole mode'', \cite{Recati2006dpo, Sommer2011sti}, which generalizes the spin-dipole mode to a fermionic mixture, or by measuring related transport phenomena \cite{Sommer2011ust}. This is a topic of ongoing research in our laboratory.

\subsection{Formation and trapping of Feshbach molecules}
\label{ssec:DyKmols}

In the physics of fermionic superfluids \cite{Varenna2006book, Giorgini2008tou, Zwerger2012tbb, Strinati2018tbb}, the formation of weakly bound molecules near a Feshbach resonance plays a particularly important role. This process combines two fermions to form a composite boson. Moreover, Pauli suppression  can lead to an amazing collisional stability of bosonic Feshbach molecules composed of fermions \cite{Petrov2004wbd, Petrov2005spo, Petrov2005dmi}.
In 2003, molecule formation was demonstrated in spin mixtures of $^{40}$K \cite{Regal2003cou} and $^{6}$Li \cite{Strecker2003coa, Cubizolles2003pol, Jochim2003pgo}. In the same year, the rapid development led to the attainment of molecular BEC \cite{Jochim2003bec, Greiner2003eoa, Zwierlein2003oob} and soon to the realization of the crossover from BEC to BCS-type systems \cite{Bartenstein2004cfa, Regal2004oor, Zwierlein2004cop, Bourdel2004eso, Partridge2005mpo}.

In Fermi-Fermi mixtures of $^{6}$Li and $^{40}$K, the formation of heteronuclear dimers was demonstrated in Refs.~\cite{Voigt2009uhf, Spiegelhalder2010aop}, but the lifetime properties \cite{Naik2011fri, Jag2016lof} turned out to be insufficient to attain molecular BEC. The recently realized mixtures of $^{6}$Li-$^{53}$Cr in Florence (M.~Zaccanti, these proceedings \cite{Zaccanti2022mif}) and $^{161}$Dy-$^{40}$K in our laboratory now offer new opportunities to reach the exciting goal of molecular BEC with heteronuclear Feshbach molecules, which then as a low-entropy system in turn can serve as a starting point for realizing novel superfluid states.

In the Dy-K mixture, the formation of molecules is rather straightforward by the standard method of ramping across a Feshbach resonance, as demonstrated in many other experiments before \cite{Chin2010fri}. We have demonstrated the formation of DyK molecules in Ref.~\cite{Ye2022ool} using the 7.3-G Feshbach resonance discussed in Sec.~\ref{ssec:DyKFR}. In a further experimental step \cite{Soave2023otf}, we captured and prepared a pure sample of these molecules in an optical dipole trap at a phase-space density close to quantum degeneracy.

In our experiments \cite{Soave2023otf}, we first prepared a deeply degenerate atomic mixture as outlined in Sec.~\ref{ssec:DyKprepare}. The Feshbach ramp for molecule association started at a magnetic field about $+50\,$mG above the resonance center. Within 0.4\,ms the field was ramped down to a magnetic detuning of $-120\,$mG, where the molecular binding energy amounts to about $h \times 220\,$kHz. The conversion efficiency (number of dimers created relative to the number of minority K atoms in the atomic mixture) was close to 30\%.

\begin{figure}
\centering 
\includegraphics[width=1.0\textwidth]{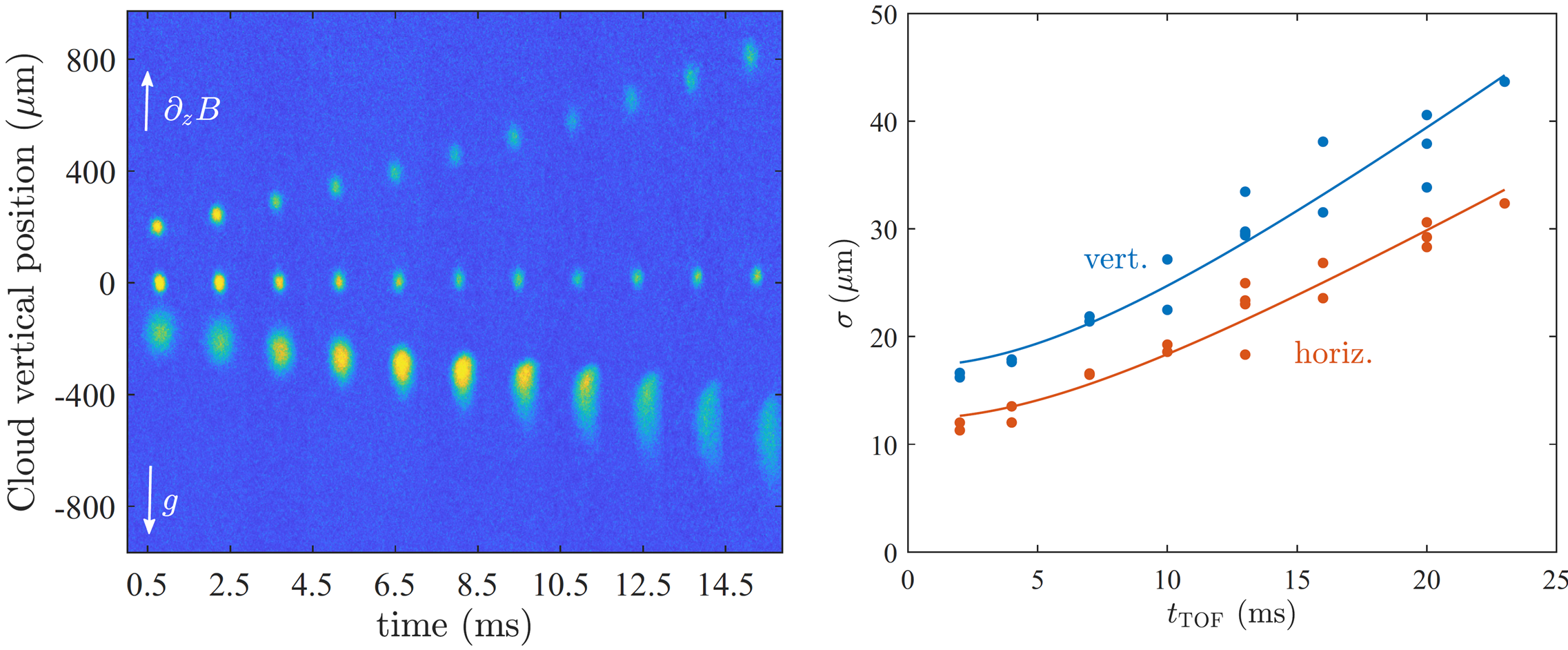}
\caption{Optically trapped DyK Feshbach molecules. The Stern-Gerlach purification process is demonstrated by the absorption images of Dy, DyK, and K (left panel), recorded in time steps of 1.5\,ms. Only the perfectly levitated molecules stay in the trap, while the Dy and K atoms leave the trap  upwards and downwards, respectively.
A time-of-flight measurement (right panel) yields the temperature of the trapped molecular cloud.
Adapted from Ref.~\cite{Soave2023otf}.}
\label{fig:DyKexpand}
\end{figure}

At this point, we adjusted the magnetic gradient to levitate the molecules at this particular magnetic field~\footnote{Note that the molecular magnetic moment, which is governed by the slope of the binding energy curve, see Fig.~\ref{fig:DyK7p3}(b), changes with the applied magnetic bias field.}, and we further reduced the power of the optical dipole trap. Under these conditions, the weak optical trap cannot hold neither the Dy nor the K atoms, which can be seen in the left panel of Fig.~\ref{fig:DyKexpand}. While the Dy atoms are overlevitated and pulled out of the trap upwards by the dominating magnetic gradient, the K atoms drop out the trap downwards. With this Stern-Gerlach technique we prepared a pure sample of about 5000 DyK molecules in the trap.

We measured the temperature of the trapped molecular cloud by time-of-flight absorption imaging, see Fig.~\ref{fig:DyKexpand} (right panel). From a fit to the horizontal expansion we obtained 45(3)\,nK, which we believe to be the true temperature~\footnote{The vertical expansion contains additional excitations caused by the rapid changes of vertical magnetic forces during the ramping process.}. 
With a mean trap frequency of $\bar{\omega}/2\pi = 28\,$Hz, we calculate a phase-space density of $\sim$$0.14$, about one order of magnitude away from degeneracy. The conditions near quantum degeneracy represent an excellent starting point for further experiments. Comparable values of the phase-space density have been reached recently in a system of NaCs Bose-Bose Feshbach molecules \cite{Lam2022hps}. While BEC has not yet been reached with bosonic heteronuclear Feshbach
samples, quantum degeneracy has been obtained in fermionic
heteronuclear systems of KRb \cite{Demarco2019adf} and NaK \cite{Duda2023tfa}.

In our experiments \cite{Soave2023otf}, we also investigated the lifetime of the pure molecular sample. The decay curves, recorded for variable hold times of the molecules in the trap, showed a purely exponential behavior without any significant contribution from inelastic two- or three-body collisions. We identified trap-light induced losses as the main cause of decay, limiting the lifetimes in our shallow 1064-nm optical dipole trap to typically 20\,ms. We also found that the lifetime increases closer to resonance. These observations suggest a light-induced one-body process in which the closed-channel fraction of the Feshbach dimer couples to molecular states in the excited-state manifold. 

This light-induced loss process, although observed also in bi-alkali systems \cite{Koeppinger2014poo, Chotia2012lld, Zhang2020fas}, seems to be a particular problem in the extrordinarily dense spectrum of molecular states in heteronuclear systems containing magnetic lanthanide atoms. In preliminary experiments we observed a fivefold reduction of losses by using trap light further in the infrared at 1547\,nm. Solving the lifetime issue by choosing the optimum wavelength for the trap light will be crucial for further progress in the experiments towards novel superfluids.

\section{Concluding remarks}

Already the two case studies presented in these lecture notes highlight the vast range of possibilities offered by quantum mixtures for future experiments. There is even much more of interesting physics to be explored with these and many other systems, as the various contributions to these proceedings impressively demonstrate. 

In our Li-K experiments, we have seen that quantum impurity physics is very rich. We could answer central questions about the properties of Fermi polarons, regarding their energy, lifetime, formation dynamics, and also their mutual interaction properties.
There are still many open fundamental questions concerning the limits of the theoretical description in terms of Fermi liquid theory, the effect of impurity motion, and the prospects of new phenomena in species-selective trapping potentials.

In our Dy-K experiments, we have demonstrated how to tame a new, unexplored mixture with exotic interaction properties. Having identified Feshbach resonances, we gained interaction control, entered the regime of strong interactions, observed hydrodynamic behavior, and finally realized a trapped molecular cloud close to quantum degeneracy. These are important steps towards the experimental realization of novel superfluids with mass imbalance.

The research field of quantum mixtures offers many exciting research opportunities. When designing a new experiment one needs to choose very carefully the best mixture with the optimum properties for the given purpose. Therefore the development of new experiments will go hand in hand with gaining or refining the understanding of the properties of particular mixtures, for which the ultracold community has developed a large tool-box of different methods.

\acknowledgments

We warmly thank all former and present team members in our fermion mixture labs (Li-K and Dy-K) as well as all our theory collaborators for their important contributions over many years to the work reported here. We also acknowledge financial support by the Austrian Science Fund FWF in various funding programs and, most recently, by the European Research Council
(ERC) under the European Union’s Horizon 2020 research
and innovation programme (Grant Agreement No. 101020438 - SuperCoolMix).

For many stimulating discussions at the Varenna Summer School we are particularly grateful to \NAME{M.~Zaccanti, D.~Petrov, M.~Parish, J.~Levinsen, C.~Salomon \atque J.~Dalibard}.

\bibliographystyle{varenna}

\end{document}